\documentclass[11pt, a4paper]{article}
\usepackage[utf8]{inputenc}
\usepackage[dvipsnames]{xcolor}
\usepackage{mathtools,amsmath,amsfonts,amssymb,amsbsy,jheppub,graphicx,hyperref}
\usepackage{bbm}
\usepackage{braket}

\def\be{\begin{equation}}
\def\ee{\end{equation}}

\preprint{\texttt{IFT-UAM/CSIC-26-113}}

\title{Quantum chaos and late-time
equipartition of symmetry-resolved Krylov complexity}

\author[a]{Jayashish Das,\href{https://orcid.org/0009-0009-8199-8901}
{\raisebox{2pt}{\includegraphics[scale=0.04]{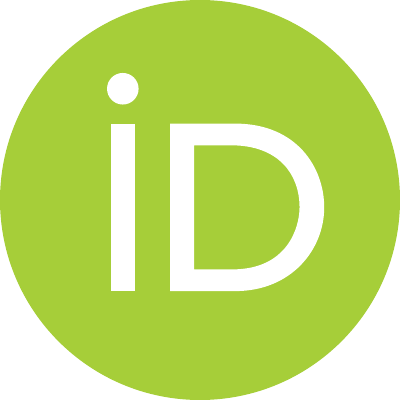}}}}

\author[b]{Suman Das,\href{https://orcid.org/0000-0002-0053-3187}
{\raisebox{2pt}{\includegraphics[scale=0.04]{orcidid.pdf}}}}
\author[b]{Juan F. Pedraza\href{https://orcid.org/0000-0002-1426-642X}{\raisebox{2pt}{\includegraphics[scale=0.04]{orcidid.pdf}}}}

\author[b,c]{and Le-Chen Qu\href{https://orcid.org/0000-0002-9712-7299}
{\raisebox{2pt}{\includegraphics[scale=0.04]{orcidid.pdf}}}}

\emailAdd{jayashish.das@saha.ac.in}
\emailAdd{suman.das@ift.csic.es}
\emailAdd{j.pedraza@csic.es}
\emailAdd{lechen.qu@ift.csic.es}

\affiliation[a]{Theory Division, Saha Institute of Nuclear Physics, A CI of Homi Bhabha National Institute, 1/AF,
Bidhannagar, Kolkata 700064, India.}

\affiliation[b]{Instituto de F\'{i}sica Te\'{o}rica UAM/CSIC, Calle Nicol\'{a}s Cabrera 13-15, Madrid, E-28049, Spain.}

\affiliation[c]{Departamento de F\'isica Te\'orica, Universidad Aut{\'o}noma de Madrid, 28049 Madrid, Spain}

\abstract{We study symmetry-resolved Krylov complexity in finite-dimensional chaotic quantum many-body systems. When both the Hamiltonian and the initial operator commute with a conserved charge, the operator dynamics decomposes into independent symmetry sectors, each with its own Krylov chain. We show that, after saturation, the unresolved Krylov complexity is additive over symmetry sectors. In the absence of additional Liouvillian degeneracies, the late-time contribution of a sector with Hilbert-space dimension $d_q$ is controlled by $d_q(d_q-1)$, leading to a dimension-weighted equipartition that approaches the simple large-sector scaling $d_q^2/\sum_{q'}d_{q'}^2$. This late-time rule differs from the early-time weighted-average discussed in the literature and is governed instead by the dimensions of the accessible operator spaces. We support the analytic prediction with numerical studies of the real and complex SYK models, a chaotic bosonic spin model, and the mixed-field Ising chain. Our results show that resolving exact symmetries is essential for interpreting the saturation value of Krylov complexity as a diagnostic of chaotic operator growth.}

\begin{document}

\maketitle

\raggedbottom

\section{Introduction}\label{sec:intro}

Understanding chaos remains a major challenge in modern physics, both theoretically and experimentally, particularly in light of recent advances in the controlled simulation of quantum many-body systems. While chaos is ubiquitous in classical dynamics, its quantum counterpart is more subtle and still lacks a complete characterization. Over the years, a variety of diagnostics have been developed to probe quantum-chaotic behavior, ranging from traditional spectral measures such as the level-spacing distribution \cite{Wigner:1951,Dyson_1962} to more recent probes such as the spectral form factor \cite{Cotler:2016fpe}.

Krylov complexity has emerged in recent years as a practical measure of operator growth under Hamiltonian evolution. It quantifies the extent to which an operator spreads in the Krylov basis generated by successive commutators with the Hamiltonian. The idea was introduced in \cite{Parker:2018yvk} (see also \cite{Barbon:2019wsy,Avdoshkin:2019trj,Rabinovici:2020ryf,Jian:2020qpp,Dymarsky:2019elm,Dymarsky:2021bjq,Rabinovici:2022beu,Hornedal:2022pkc,He:2022ryk,Kundu:2023hbk,Caputa:2023vyr,He:2024hkw,He:2024xjp,Caputa:2024vrn,Craps:2024suj,Bhattacharya:2024uxx,Aguilar-Gutierrez:2024nau,Basu:2025ubf,Craps:2025kub,He:2025guu,Caputa:2025mii,Demulder:2025uda,Imani:2025etp,DeRo:2026mlc,Muck:2022xfc,Kar:2021nbm,Lunt:2025dcc,Kotta:2026utn,Nastase:2026lhz,Murugan:2026rfa} and the recent reviews \cite{Nandy:2024evd,Rabinovici:2025otw}), where it was argued that Krylov complexity exhibits exponential growth in chaotic systems.\footnote{A closely related construction applies to states evolving in the Schr\"odinger picture, where the Krylov basis is generated by repeated action of the Hamiltonian on an initial state. The associated Krylov state complexity, often referred to as `spread complexity,' was introduced in Ref.~\cite{Balasubramanian:2022tpr} and it has been studied in a variety of gravitational and quantum many-body systems; see, e.g., Refs.~\cite{Erdmenger:2023wjg,Craps:2023ivc,Huh:2023jxt,Caputa:2024sux,Zhai:2024tkz,Nandy:2024mml,Li:2024ljz,Balasubramanian:2024ghv,Bhattacharya:2024szw,Bhattacharya:2024hto,Camargo:2024deu,Baggioli:2024wbz,Huh:2024ytz,Jeong:2024jjn,Baggioli:2025ohh,Baggioli:2025knt,Fu:2025kkh,Zhai:2025abc,Caputa:2025dep,Caputa:2025ozd,Takahashi:2025iol,Miyaji:2025ucp,Jeong:2026iac,Alishahiha:2026fnu,Chowdhury:2026fjb,Li:2026jxx,Nunez:2026kwr,Muck:2024fpb,Alishahiha:2024vbf,Balasubramanian:2025xkj,Balasubramanian:2022dnj,Qu:2025lgo,Fatemiabhari:2025usn,Fatemiabhari:2026goj,Begines:2026fnx,Basu:2026gvl,Nunez:2026vhw,Pedraza:2026zji,Qu:2026dmv,Erdmenger:2026iga,Forste:2025gng} and the review~\cite{Jeong:2026gdc}.} However, it was subsequently shown in \cite{Dymarsky:2019elm,Dymarsky:2021bjq} that such exponential growth is in fact a universal feature of continuum quantum field theories. Consequently, exponential growth alone cannot serve as a reliable diagnostic of quantum chaos in such systems. A modified prescription aimed at circumventing this ambiguity was proposed in \cite{Kundu:2023hbk}.

For finite-dimensional systems, the situation is clearer. Beginning with the work of \cite{Rabinovici:2020ryf} and subsequently \cite{Rabinovici:2022beu,Basu:2025ubf}, it is now understood that Krylov complexity in chaotic many-body systems typically exhibits an initial exponential growth up to the scrambling time, $t\sim \log S$, followed by a linear growth regime up to the Heisenberg time, $t\sim e^{S}$, after which it saturates. Throughout this discussion, we remain well below the Poincar\'e recurrence time, of order $t\sim e^{e^{S}}$. In practice, identifying the early-time exponential regime can be difficult. The late-time saturation value therefore provides a more robust diagnostic for distinguishing between integrable and chaotic dynamics. In chaotic systems, this value is typically much larger than in integrable systems, reflecting the fact that a generic local operator explores a larger portion of its dynamically accessible operator space.

Symmetries provide a natural organizing principle for quantum many-body dynamics. In the presence of a conserved charge, the Hilbert space decomposes into distinct symmetry sectors labeled by the corresponding quantum numbers. Physical observables can consequently acquire a nontrivial sector dependence, and failing to resolve these sectors may obscure important information. A familiar example is provided by spectral statistics: random-matrix behavior is expected to emerge within individual symmetry sectors, whereas mixing different sectors can introduce additional degeneracies and distort the level-spacing distribution. Symmetry resolution therefore provides a more refined characterization of quantum dynamics by separating contributions from distinct invariant subspaces.

A prominent example is symmetry-resolved entanglement. The reduced density matrix of a subsystem can be decomposed into fixed-charge sectors, allowing one to ask how the total entanglement is distributed among them \cite{laflorencie14spin,PhysRevLett.120.200602,xavier2018,lukin19exp}. In many equilibrium and nonequilibrium settings, the leading contribution to the symmetry-resolved entanglement entropy becomes independent of the charge sector in an appropriate scaling regime, a phenomenon known as entanglement equipartition \cite{xavier2018}. This picture extends naturally to dynamical settings, such as quantum quenches or driven systems, where equipartition is associated with the late-time or large-subsystem regime, while finite-size, finite-time, or protocol-dependent effects can lead to visible violations \cite{Ares:2026vjt}.

It is natural to ask whether an analogous sector-wise structure arises for operator growth. If the Hamiltonian and the seed operator commute with a common symmetry, both decompose into symmetry blocks, and a Krylov complexity can be defined independently within each sector. This leads to the notion of symmetry-resolved Krylov complexity, formalized in \cite{Caputa:2025mii} (see also \cite{Kotta:2026utn,Nastase:2026lhz, Murugan:2026wss}).\footnote{Related ideas appeared earlier in \cite{Rabinovici:2020ryf}, where Krylov complexity was computed within a particular symmetry sector.} These works established that, at early times, the full Krylov complexity is related to a weighted average of the symmetry-resolved complexities, with weights determined by the projection of the initial operator onto the different sectors. This relation is not an equipartition statement: it follows from the short-time Krylov expansion and retains explicit information about the initial operator. At late times, however, no analogous relation between the full and symmetry-resolved complexities is known. This raises the question of whether the early-time weighted relation persists after saturation or gives way to a different sector-wise structure.

In this work, we address this question for finite-dimensional chaotic systems. Under generic spectral assumptions and effective late-time delocalization of the Krylov wavefunction over the accessible Krylov space, we find that in the saturation regime,
\begin{equation}
t_{\rm sat}\lesssim t\ll t_{\rm rec},
\end{equation}
the full Krylov complexity is additive over symmetry sectors,
\begin{equation}
C_K(t)
\simeq
\sum_q C_K^{(q)}(t).
\end{equation}
In the absence of additional Liouvillian degeneracies, the corresponding fractional complexity satisfies
\begin{equation}
\frac{C_K^{(q)}}{C_K}
\simeq
\frac{d_q(d_q-1)}
{\sum_{q'}d_{q'}(d_{q'}-1)}
\simeq
\frac{d_q^2}{\sum_{q'}d_{q'}^2},
\end{equation}
where $d_q$ denotes the Hilbert-space dimension of sector $q$, and the second expression applies for large sectors. We refer to this behavior as \emph{dimension-weighted equipartition}: sectors do not contribute equally in general, but rather according to the dimensions of their dynamically accessible operator spaces.

This late-time relation is qualitatively distinct from its early-time counterpart. In particular, the initial sector weights drop out after saturation, and the fractional complexities are controlled instead by the accessible Krylov-space dimensions. This provides a simple generic relation between the full and symmetry-resolved Krylov complexities in chaotic systems and highlights the importance of resolving exact symmetries. Indeed, one might naively expect the saturation value in a chaotic system to scale as
\begin{equation}
C_K^{\rm sat}\sim D^2=\left(\sum_q d_q\right)^2,
\end{equation}
where $D$ is the total Hilbert-space dimension. Instead, under the assumptions above, we find the large-sector scaling
\begin{equation}
C_K^{\rm sat}\sim \sum_q d_q^2,
\end{equation}
which can be parametrically smaller than $D^2$. Symmetry resolution is therefore essential for correctly interpreting the late-time saturation of Krylov complexity as a diagnostic of chaotic operator growth.

The paper is organized as follows. In section~\ref{sec:definitions}, we introduce the setup and define symmetry-resolved Krylov complexity. In section~\ref{sec:late-time-saturation}, we derive the late-time relation between the full and symmetry-resolved complexities and discuss its implications. In section~\ref{sec:algorithm}, we present the numerical algorithm used to compute Krylov complexity. In section~\ref{sec:numerics}, we analyze several models and compare the numerical results with our analytic predictions. We conclude in section~\ref{sec:discussion} with a summary of our findings and a discussion of future directions. Some details are relegated to the appendices. In appendix~\ref{app:wavefn_property}, we examine the late-time saturation of the Krylov wavefunction, while in appendix~\ref{app:Krylov_integrable} we investigate symmetry-resolved Krylov complexity in integrable systems.


\section{Preliminaries}\label{sec:definitions}

We begin with a brief review of the Krylov-space framework used throughout
this work to quantify operator growth under Heisenberg evolution. Further
technical details can be found in~\cite{Parker:2018yvk},
while pedagogical introductions are provided in the recent
reviews~\cite{Nandy:2024evd,Rabinovici:2025otw}.

\subsection{Krylov complexity}
Krylov complexity characterizes operator growth under Heisenberg evolution by
quantifying how far an initially simple operator spreads in a basis generated
by repeated commutation with the Hamiltonian. Consider a finite-dimensional
quantum system with a Hermitian Hamiltonian $H$ and a Hermitian seed operator
$O$. The Heisenberg evolution of $O$ is
\begin{equation}
    O(t)=e^{iHt}Oe^{-iHt}
        =e^{i\mathcal{L}t}O,
\end{equation}
where the Liouvillian superoperator is defined by
$\mathcal{L}\equiv[H,\cdot]$. Expanding about $t=0$, we obtain
\begin{equation}
    O(t)
    =
    \sum_{n=0}^{\infty}
    \frac{(it)^n}{n!}\,
    \mathcal{L}^{\,n}O.
\end{equation}
It follows that the entire time evolution is confined to the Krylov space
\begin{equation}
    \mathcal{K}_{O}
    =
    \operatorname{span}
    \left\{
    O,\mathcal{L}O,\mathcal{L}^{2}O,\ldots
    \right\}.
\end{equation}
The operators in this sequence span the dynamically accessible operator
space, but they are generally neither normalized nor mutually orthogonal. The
Lanczos algorithm transforms this sequence into an orthonormal Krylov basis
and, at the same time, brings the Liouvillian into tridiagonal form. To define orthogonality in operator space, we adopt the inner
product,
\begin{equation}\label{inner}
    (A|B)
    \equiv
    \operatorname{Tr}\!\left(A^\dagger B\right).
\end{equation}
With the inner product in
Eq.~\eqref{inner}, the Liouvillian $\mathcal{L}=[H,\cdot]$ is self-adjoint in
Liouville space. The Lanczos construction begins with the normalized operator state
\begin{equation}
    |O_0)
    =
    \frac{|O)}{\sqrt{(O|O)}},
    \qquad
    |O_{-1})=0,
    \qquad
    b_0=0.
\end{equation}
Suppose that the first $n+1$ Krylov basis vectors have already been
constructed. Acting with $\mathcal{L}$ on $|O_n)$ generates a candidate for the
next basis vector. For the Hermitian seed operator, the diagonal element vanishes,
\begin{equation}
    (O_n|\mathcal{L}|O_n)=0.
\end{equation}
Consequently, one only needs to subtract the component parallel to the
preceding Krylov vector:
\begin{align}
    |A_{n+1})
    &=
    \mathcal{L}|O_n)
    -b_n|O_{n-1}),
    \nonumber\\
    b_{n+1}
    &=
    \sqrt{(A_{n+1}|A_{n+1})},
    \nonumber\\
    |O_{n+1})
    &=
    \frac{|A_{n+1})}{b_{n+1}}.
\end{align}
The recursion terminates when $b_{n+1}=0$, since at that point the action of
the Liouvillian generates no new linearly independent operator direction. The
resulting Krylov basis is orthonormal,
$(O_m|O_n)=\delta_{mn}$, and the Liouvillian acts as
\begin{equation}
    \mathcal{L}|O_n)
    =
    b_{n+1}|O_{n+1})
    +b_n|O_{n-1}).
\end{equation}
The positive numbers $b_n$ are the Lanczos coefficients. This relation maps
the Liouvillian dynamics onto nearest-neighbor hopping along a
one-dimensional Krylov chain: $b_n$ are the hopping amplitudes, and the
vanishing diagonal matrix elements imply the absence of on-site terms. Let $K=\dim\mathcal{K}_{O}$ denote the dimension of the Krylov space. The
normalized time-evolved operator can be expanded in the Krylov basis as
\begin{equation}
    |O(t))
    =
    \sum_{n=0}^{K-1}
    i^n\phi_n(t)\,|O_n),
    \qquad
    \sum_{n=0}^{K-1}|\phi_n(t)|^2=1.
\end{equation}
Here, $|\phi_n(t)|^2$ gives the probability of finding the operator
wavefunction at site $n$ of the Krylov chain. Introducing the Krylov position
operator,
\begin{equation}
    \widehat{C}_{K}
    =
    \sum_{n=0}^{K-1}n\,|O_n)(O_n|,
\end{equation}
we define the Krylov complexity as
\begin{equation}
    C_K(t)
    =
    (O(t)|\widehat{C}_{K}|O(t))
    =
    \sum_{n=0}^{K-1}
    n\,|\phi_n(t)|^2.
\end{equation}
Thus, $C_K(t)$ is the mean position of the operator wavefunction along the
Krylov chain. A larger value of $C_K(t)$ indicates that the evolving operator
has acquired significant support on higher-order nested commutators and has
therefore explored a larger fraction of its dynamically accessible operator
space. 
\subsection{Symmetry-resolved Krylov complexity}
We now turn to systems endowed with a conserved symmetry generated by a
Hermitian operator $P$. We assume that the Hamiltonian is invariant under this
symmetry and restrict our attention to a Hermitian seed operator that is
itself symmetry invariant:
\begin{equation}
    [H,P]=0,
    \qquad
    [O(0),P]=0.
\end{equation}
At the same time, we require
\begin{equation}
    [H,O(0)]\neq0,
\end{equation}
so that the operator undergoes nontrivial Heisenberg evolution. Since $P$
commutes with $H$, symmetry invariance is preserved in time:
\begin{equation}
    [O(t),P]
    =
    e^{iHt}[O(0),P]e^{-iHt}
    =
    0.
\end{equation}
Let $\lambda_q$ denote the distinct eigenvalues of $P$, and let
$\mathcal{H}_q$ be the corresponding eigenspaces, with
$d_q=\dim\mathcal{H}_q$. The Hilbert space therefore decomposes into symmetry
sectors as
\begin{equation}
    \mathcal{H}
    =
    \bigoplus_q\mathcal{H}_q,
    \qquad
    P
    =
    \bigoplus_q
    \lambda_q\,\mathbb{I}_{d_q}.
\end{equation}
The projector onto $\mathcal{H}_q$ can be written as
\begin{equation}
    \Pi_q
    =
    \prod_{q'\neq q}
    \frac{P-\lambda_{q'}\mathbb{I}}
         {\lambda_q-\lambda_{q'}},
    \qquad
    \Pi_q\Pi_{q'}
    =
    \delta_{qq'}\Pi_q,
    \qquad
    \sum_q\Pi_q
    =
    \mathbb{I}.
\end{equation}
The commutation relations above imply that neither the Hamiltonian nor the
evolving operator mixes different symmetry sectors. Defining their restrictions
to $\mathcal{H}_q$ by
\begin{equation}
    H_q
    =
    \Pi_qH\Pi_q,
    \qquad
    O_q(t)
    =
    \Pi_qO(t)\Pi_q,
\end{equation}
we obtain the block decompositions
\begin{equation}
    H
    =
    \bigoplus_qH_q,
    \qquad
    O(t)
    =
    \bigoplus_qO_q(t).
\end{equation}
Moreover, because $[H,\Pi_q]=0$, each block evolves independently within its
own symmetry sector:
\begin{equation}
    O_q(t)
    =
    e^{iH_qt}O_q(0)e^{-iH_qt}.
\end{equation}
This block decomposition is the starting point for resolving operator growth
with respect to the conserved symmetry. As a simple illustration, consider a $\mathbb{Z}_2$ symmetry satisfying
$P^2=\mathbb{I}$. Its eigenvalues are $\lambda_\pm=\pm1$, and the corresponding
projectors are
\begin{equation}
    \Pi_\pm
    =
    \frac{1}{2}
    \left(\mathbb{I}\pm P\right).
\end{equation}
The Hilbert space decomposes as
\begin{equation}
    \mathcal{H}
    =
    \mathcal{H}_+
    \oplus
    \mathcal{H}_-,
\end{equation}
and, in the eigenbasis of $P$, the Hamiltonian and the invariant operator take
the block-diagonal form
\begin{equation}
    H
    =
    \begin{pmatrix}
        H_+ & 0 \\
        0   & H_-
    \end{pmatrix},
    \qquad
    O(t)
    =
    \begin{pmatrix}
        O_+(t) & 0 \\
        0      & O_-(t)
    \end{pmatrix}.
\end{equation}
In practice, this basis may be obtained by diagonalizing
$P=UDU^\dagger$ and transforming
$H\mapsto U^\dagger HU$ and $O\mapsto U^\dagger OU$. The commutation of $H$
and $O$ with $P$ guarantees that both transformed operators are block
diagonal, with blocks organized according to the repeated eigenvalues of $D$.
More precisely, if an eigenvalue $\lambda_q$ appears $d_q$ times in $D$, it
defines a $d_q$-fold degenerate eigenspace $\mathcal{H}_q$ of $P$. The
corresponding blocks $H_q$ and $O_q$ act within this degenerate eigenspace and
therefore have dimensions $d_q\times d_q$. Having isolated the symmetry blocks, we apply the Krylov construction to each
of them separately \cite{Caputa:2025mii}. In sector $q$, the relevant
Liouvillian is
\begin{equation}
    \mathcal{L}_q
    \equiv
    [H_q,\cdot],
\end{equation}
and the fixed-sector Krylov basis
$\{|O_n^{(q)})\}_{n=0}^{K_q-1}$ is generated from the normalized block
$|O_0^{(q)})$. Expanding the normalized evolving block as
\begin{equation}
    |O_q(t))
    =
    \sum_{n=0}^{K_q-1}
    i^n\phi_n^{(q)}(t)\,
    |O_n^{(q)}),
    \qquad
    \sum_{n=0}^{K_q-1}
    \left|\phi_n^{(q)}(t)\right|^2
    =
    1,
\end{equation}
we define the symmetry-resolved Krylov complexity by
\begin{equation}
    C_K^{(q)}(t)
    =
    \sum_{n=0}^{K_q-1}
    n\,
    \left|\phi_n^{(q)}(t)\right|^2.
\end{equation}
It measures the growth of the single block $O_q(t)$ and, by construction,
does not include the other symmetry sectors. To compare the resolved quantities with the evolution of the full operator,
we introduce the sector weights
\begin{equation}
    p_q
    =
    \frac{(O_q|O_q)}{(O|O)},
    \qquad
    \sum_qp_q
    =
    1.
\end{equation}
If the full operator state and each block state are normalized separately,
their relation can be written as
\begin{equation}
    |O(t))
    =
    \sum_q
    \sqrt{p_q}\,
    |O_q(t)).
\end{equation}
This motivates the sector-averaged Krylov complexity
\begin{equation}
    \overline{C}(t)
    =
    \sum_q
    p_q\,C_K^{(q)}(t).
\end{equation}
It is important to distinguish the exact block decomposition of the operator
from a decomposition of its Krylov complexity. Although $O(t)$ is the direct
sum of the blocks $O_q(t)$, the full Krylov basis is obtained by
orthonormalizing all nested commutators simultaneously, whereas the
fixed-sector bases are constructed by independent orthonormalizations within
each block. Consequently, the full Krylov basis does not, in general,
decompose into fixed-sector Krylov vectors with the same Krylov index.
Accordingly, $\overline{C}(t)$ captures the leading early-time growth of the
full complexity, but it need not coincide with $C_K(t)$ at arbitrary times
\cite{Caputa:2025mii}. Establishing their relation in the late-time regime is
one of the central objectives of the present work. For chaotic finite-dimensional systems, we find that this relation simplifies
considerably at late times. At times of order $t\sim e^S$, when the Krylov
complexities have reached their saturation plateaux, the full Krylov
complexity is equal to the sum of the symmetry-resolved contributions:
\begin{equation}
    \lim_{t\to\infty}C_K(t)
    =
    \sum_q
    \lim_{t\to\infty}C_K^{(q)}(t).
    \label{eq:late-time-additivity}
\end{equation}
Here, the late-time limit is understood as the saturation regime prior to
Poincar\'e recurrences. In the next section, we establish
Eq.~\eqref{eq:late-time-additivity} by determining the dimensions of the full
and symmetry-resolved Krylov spaces.

\section{Late-time saturation of symmetry-resolved Krylov complexity}
\label{sec:late-time-saturation}

In the previous section, we defined the Krylov complexity of the full
invariant operator and the corresponding symmetry-resolved complexities of
its blocks. We also emphasized that their relation is nontrivial at generic
times because the unresolved and fixed-sector Krylov bases are obtained
through different orthonormalization procedures. In this section, we show
that a simple additive relation nevertheless emerges in the late-time
saturation regime of a finite-dimensional chaotic system. The argument
proceeds in two steps. We first determine the dimensions of the accessible
Krylov spaces, allowing for possible Liouvillian degeneracies and vanishing
seed overlaps, and subsequently specialize to the generic chaotic case. We
then use the late-time delocalization of the Krylov wavefunctions to evaluate
their saturation values.

\subsection{Krylov dimension within a symmetry sector}

Consider a symmetry sector $\mathcal{H}_q$ of dimension $d_q$, and let
\begin{equation}
    H_q|E_{a,q}\rangle
    =
    E_{a,q}|E_{a,q}\rangle,
    \qquad
    a=1,\ldots,d_q.
\end{equation}
The matrix units
\begin{equation}
    |\omega_{ab,q})
    \equiv
    |E_{a,q}\rangle\langle E_{b,q}|
\end{equation}
form an eigenbasis of the fixed-sector Liouvillian
$\mathcal{L}_q=[H_q,\cdot]$:
\begin{equation}
    \mathcal{L}_q|\omega_{ab,q})
    =
    \omega_{ab,q}|\omega_{ab,q}),
    \qquad
    \omega_{ab,q}
    \equiv
    E_{a,q}-E_{b,q}.
    \label{eq:sector-liouvillian-spectrum}
\end{equation}
The block of the seed operator in this sector can therefore be expanded as
\begin{equation}
    |O_q)
    =
    \sum_{a=1}^{d_q}
    O_{aa,q}|\omega_{aa,q})
    +
    \sum_{\substack{a,b=1\\a\neq b}}^{d_q}
    O_{ab,q}|\omega_{ab,q}).
    \label{eq:sector-operator-expansion}
\end{equation}
The coefficient \(O_{ab,q}\) is the matrix element of the seed block \(O_q\)
in the energy eigenbasis of the symmetry sector \(q\):
\begin{equation}
    O_{ab,q}
    \equiv
    \langle E_{a,q}|O_q|E_{b,q}\rangle.
    \label{eq:seed-matrix-element}
\end{equation}
It determines the component of the seed along the Liouvillian eigenoperator
\begin{equation}
    \lvert\omega_{ab,q})
    =
    \lvert E_{a,q}\rangle
    \langle E_{b,q}\rvert.
\end{equation}
Repeated action of the Liouvillian gives
\begin{align}
    \mathcal{L}_q^n|O_q)
    ={}
    \delta_{n0}
    \sum_{a=1}^{d_q}
    O_{aa,q}|\omega_{aa,q})
   +
    \sum_{\substack{a,b=1\\a\neq b}}^{d_q}
    O_{ab,q}\,
    \omega_{ab,q}^{\,n}
    |\omega_{ab,q}).
    \label{eq:krylov-sector}
\end{align}
The factor \(\delta_{n0}\) in Eq.~\eqref{eq:krylov-sector} follows from the fact that all diagonal matrix units belong to the zero-frequency eigenspace of the Liouvillian.

To account simultaneously for Liouvillian degeneracies and vanishing seed overlaps, and following the standard spectral-function formulation of operator Krylov dynamics~\cite{Parker:2018yvk,Nandy:2024evd,hsvm-w849}, we
first define the component of the sector seed projected onto the
Liouvillian eigenspace with distinct frequency \(\omega\) as
\begin{equation}
    |O_{\omega,q})
    \equiv
    \sum_{\substack{1\leq a,b\leq d_q\\
                    \omega_{ab,q}=\omega}}
    O_{ab,q}\,|\omega_{ab,q}).
    \label{eq:frequency-projected-seed}
\end{equation}
Since every matrix unit entering this sum has the same Liouvillian eigenvalue \(\omega\), the projected component satisfies
\begin{equation}
    \mathcal{L}_q|O_{\omega,q})
    =
    \omega|O_{\omega,q}).
    \label{eq:frequency-projected-seed-eigenvalue}
\end{equation}
The sector-resolved spectral weight associated with the distinct frequency \(\omega\) is then defined as the squared Hilbert--Schmidt norm of this projected component:
\begin{align}
    W_q(\omega)
    &\equiv
    (O_{\omega,q}|O_{\omega,q})
    \nonumber=
    \sum_{\substack{1\leq a,b\leq d_q\\
                    \omega_{ab,q}=\omega}}
    \left|O_{ab,q}\right|^2 ,
    \label{eq:sector-spectral-weight}
\end{align}
and introduce the corresponding seed-accessible frequency set
\begin{equation}
    \Omega_q
    \equiv
    \left\{
        \text{distinct}\ \omega\,\middle|\,W_q(\omega)>0
    \right\}.
    \label{eq:accessible-frequency-set}
\end{equation}
Since $\mathcal{L}_q$ is self-adjoint in Liouville space, the cyclic subspace generated by a nonvanishing seed $|O_q)$ contains one independent direction for each distinct frequency with nonzero spectral weight. The exact fixed-sector Krylov dimension is therefore
\begin{equation}
    K_q=\left|\Omega_q\right|.
    \label{eq:exact-sector-Kq}
\end{equation}
This expression makes the dependence on the seed operator explicit. It is
also useful to introduce the complete difference spectrum of the Hamiltonian
block,
\begin{equation}
    \Omega_{H_q}
    \equiv
    \left\{
        \omega_{ab,q}
        \,\middle|\,
        a,b=1,\ldots,d_q
    \right\},
    \qquad
    K_{H_q}\equiv\left|\Omega_{H_q}\right|.
    \label{eq:block-difference-spectrum}
\end{equation}
Since $\Omega_q\subseteq\Omega_{H_q}$, the corresponding dimensions satisfy
\begin{equation}
    K_q
    \leq
    K_{H_q}
    \leq
    d_q(d_q-1)+1.
    \label{eq:Kq-KHq-bound}
\end{equation}
The first inequality is saturated when the seed has nonzero overlap with every distinct Liouvillian eigenspace. The second is saturated when the energy spectrum is non-degenerate and all nonzero energy differences are distinct.\footnote{A non-degenerate energy spectrum does not, by itself, imply a non-degenerate difference spectrum.}

\subsection{Dimension of the unresolved Krylov space}
\label{subsec:unresolved-Krylov-dimension}

We next consider the full invariant seed $|O)=\bigoplus_q|O_q)$. Restricting to sectors for which $O_q\neq0$, the accessible frequency set of the full
Liouvillian is
\begin{equation}
    \Omega
    =
    \bigcup_q\Omega_q,
    \qquad
    K
    =
    \left|\Omega\right|
    =
    \left|\bigcup_q\Omega_q\right|.
    \label{eq:unresolved-frequency-set}
\end{equation}
Indeed, when the same frequency occurs in several sectors, with nonzero seed overlap, the repeated action of the full Liouvillian generates only the fixed direct sum combination selected by the seed. The unresolved Krylov dimension is therefore determined by the cardinality of the union of the fixed-sector frequency sets, rather than by the sum of their
cardinalities. In general,
\begin{equation}
    K\leq\sum_q K_q.
    \label{eq:general-union-bound}
\end{equation}
If $W_q(0)>0$ in every nonempty sector, all the sets $\Omega_q$ contain the
common zero frequency, and hence
\begin{equation}
    K
    \leq
    1+\sum_q\left(K_q-1\right).
    \label{eq:general-unresolved-bound}
\end{equation}
Equality holds if and only if the nonzero accessible frequency sets $\Omega_q\setminus\{0\}$ are mutually disjoint.

For a generic chaotic Hamiltonian, after resolving all exact symmetries, and for a generic symmetry-invariant seed operator, one expects nonzero seed overlap with every symmetry-allowed Liouvillian frequency. Moreover, apart from the universally degenerate zero-frequency component, the energy differences are generically nondegenerate both within and across symmetry sectors. Under these conditions, the unresolved Krylov dimension saturates its maximal  bound. Assuming $W_q(0)>0$ in every sector, one then has
\begin{equation}
    K_q=d_q(d_q-1)+1,
    \qquad
    \Omega_q\cap\Omega_{q'}=\{0\},
    \qquad
    q\neq q'.
    \label{eq:generic-sector-frequencies}
\end{equation}
Consequently,
\begin{equation}
    K
    =
    1+\sum_q d_q(d_q-1)
    =
    1+\sum_q\left(K_q-1\right)
    =
    \sum_qK_q-N_{\rm sec}+1,
    \label{eq:generic-unresolved-K}
\end{equation}
where $N_{\rm sec}$ denotes the number of nonempty symmetry sectors.\footnote{An important exception occurs in the real SYK model whenever $N\equiv 2 \pmod 4$. In this case, the antiunitary particle--hole symmetry exchanges the two fermion-parity sectors and renders them isospectral. Their difference spectra therefore coincide, invalidating the assumption of cross-sector non-degeneracy and reducing the unresolved Krylov dimension. The case $N=14$ provides one such example.}

In integrable systems, degeneracies of the energy differences and selection rules associated with the seed operator can invalidate either of the generic assumptions above. Under the assumption that $W_q(0)>0$ in every sector, one may therefore have
\begin{equation}
    K_q<d_q(d_q-1)+1
    \qquad\text{and/or}\qquad
    K<1+\sum_q\left(K_q-1\right).
    \label{eq:integrable-Krylov-dimensions}
\end{equation}
This is illustrated in detail for integrable models in
appendix~\ref{app:Krylov_integrable}.

\subsection{Late-time saturation and additivity}\label{subsec:late_time_Krylov}
We now translate the dimension counting into a statement about late-time
Krylov complexity. For chaotic finite-dimensional systems, we assume that,
after the saturation time $t_{\mathrm{sat}}$, the Krylov wavefunction is
effectively delocalized over the accessible Krylov chain
\cite{Rabinovici:2020ryf,Balasubramanian:2023kwd}. In a fixed sector this gives
\begin{equation}
    \left|\phi_n^{(q)}(t)\right|^2
    \simeq
    \frac{1}{K_q},
    \qquad
    t\gtrsim t_{\mathrm{sat}},
    \qquad
    n=0,\ldots,K_q-1.
    \label{eq:sector-uniform-krylov-distribution}
\end{equation}
The corresponding saturation value is
\begin{align}
    C_K^{(q)}(t\gtrsim t_{\mathrm{sat}})
    \simeq
    \frac{1}{K_q}
    \sum_{n=0}^{K_q-1}n
    =
    \frac{K_q-1}{2}
    =
    \frac{d_q(d_q-1)}{2},
    \label{eq:sector-krylov-saturation}
\end{align}
where Eq.~\eqref{eq:generic-sector-frequencies} was used in the last step. Applying the same delocalization assumption to the unresolved Krylov chain
yields
\begin{align}
    C_K(t\gtrsim t_{\mathrm{sat}})
    \simeq
    \frac{K-1}{2}
    =
    \frac{1}{2}
    \sum_qd_q(d_q-1),
    \label{eq:full-krylov-saturation}
\end{align}
where we used Eq.~\eqref{eq:generic-unresolved-K}. Comparing
Eqs.~\eqref{eq:sector-krylov-saturation} and
\eqref{eq:full-krylov-saturation}, we obtain
\begin{equation}
    C_K(t\gtrsim t_{\mathrm{sat}})
    \simeq
    \sum_q
    C_K^{(q)}(t\gtrsim t_{\mathrm{sat}}).
    \label{eq:late-time-additivity-derived}
\end{equation}
Equivalently, with the late-time limit understood as the saturation plateau
before Poincar\'e recurrences,
\begin{equation}
    \lim_{t\to\infty}C_K(t)
    =
    \sum_q
    \lim_{t\to\infty}C_K^{(q)}(t),
\end{equation}
which proves the relation announced in
Eq.~\eqref{eq:late-time-additivity}. The result follows from two ingredients:
the counting of accessible Liouvillian modes and the effective late-time
delocalization of the Krylov wavefunction. It is therefore expected to receive
corrections whenever additional Liouvillian degeneracies, vanishing seed
overlaps, finite-size effects, or incomplete Krylov-space equilibration are
important.

\section{The algorithm}\label{sec:algorithm}

In this section, we briefly describe the numerical procedure used to compute the Krylov complexity. The first step is the construction of the Krylov basis associated with a Hamiltonian $H$ and an operator $O$. This is achieved by applying the Lanczos algorithm to the Liouvillian superoperator
\begin{equation}
    \mathcal{L}\equiv [H,\cdot] .
\end{equation}
The algorithm described below assumes that all coefficients $O_{ab,q}$ are
nonzero, as is the case for the chaotic models considered in the main text.
When this condition is not satisfied, we use the alternative procedure
described in appendix~\ref{app:Krylov_integrable}.

Starting from the normalized seed operator
\begin{equation}
    |O_0)=\frac{|O)}{\sqrt{(O|O)}} ,
\end{equation}
the Krylov basis vectors are generated recursively. Defining $ b_0=0, |O_{-1})=0 ,$
the recursion relations are
\begin{align}
    a_{n-1}
    &=
    (O_{n-1}|\mathcal{L}|O_{n-1}) ,
    \nonumber\\
    |A_n)
    &=
    \mathcal{L}|O_{n-1})
    -a_{n-1}|O_{n-1})
    -b_{n-1}|O_{n-2}) ,
    \nonumber\\
    b_n
    &=
    \sqrt{(A_n|A_n)} ,
    \nonumber\\
    |O_n)
    &=
    \frac{|A_n)}{b_n} , \quad a_n=(O_n| \mathcal{L} |O_n) .
\end{align}
The iteration is continued until the Krylov space is exhausted, i.e., until $(A_n|A_n)=0$. The resulting Lanczos coefficients $\{a_n,b_n\}$ encode the effective hopping amplitudes of the Krylov chain and completely characterize the operator dynamics in the Krylov basis.

Once the Krylov basis has been constructed, the time-evolved operator may be expanded as
\begin{equation}
    |O(t))
    =
    \sum_{n=0}^{K-1}
    i^n \phi_n(t)\,|O_n) ,
\end{equation}
where $\phi_n(t)$ denotes the Krylov wavefunction and $K$ is the Krylov dimension. The Krylov complexity is then given by
\begin{equation}
    C_K(t)
    =
    \sum_{n=0}^{K-1}
    n\,|\phi_n(t)|^2 .
\end{equation}
In practical implementations, the Lanczos recursion may suffer from a loss of orthogonality at large Krylov orders due to the accumulation of numerical errors. To improve numerical stability, we employ the Arnoldi iteration, which explicitly re-orthogonalizes the basis vectors at every step. We follow refs.~\cite{Bhattacharjee:2022lzy,Bhattacharya:2022gbz} and adopt their notation. Starting from
\begin{equation}
    |V_0)
    =
    \frac{|O_0)}
    {\sqrt{(O_0|O_0)}} ,
\end{equation}
the Arnoldi basis is generated through

\begin{align}
    |U_k)
    &=
    \mathcal{L}|V_{k-1}) ,
    \nonumber\\
    h_{j,k-1}
    &=
    (V_j|U_k) ,
    \nonumber\\
    |\widetilde{U}_k)
    &=
    |U_k)
    -
    \sum_{j=0}^{k-1}
    h_{j,k-1}|V_j) ,
    \nonumber\\
    h_{k,k-1}
    &=
    \sqrt{
    (\widetilde{U}_k|\widetilde{U}_k)
    } ,
    \nonumber\\
    |V_k)
    &=
    \frac{|\widetilde{U}_k)}
    {h_{k,k-1}} .
    \label{eq:arnoldi_alogo}
\end{align}
The iteration is continued until $h_{k, k-1}=0$. For Hermitian Liouvillians, the Arnoldi procedure is equivalent to the Lanczos construction up to a change of basis. The Lanczos coefficients are extracted from the Hessenberg matrix according to
\begin{equation}
    a_n=h_{n,n},
    \qquad
    b_n=h_{n,n-1}.
\end{equation}
All numerical results for the chaotic models in the main text are obtained using this Arnoldi implementation.
%

\section{Numerical studies}\label{sec:numerics}

In this section, we provide numerical evidence to test the analytic results obtained in section~\ref{sec:late-time-saturation}. For concreteness, we investigate a variety of many-body systems, including the real and complex SYK models, a bosonic spin model, and the mixed-field Ising model.

\subsection{Real SYK model}

In this section, we consider the Sachdev--Ye--Kitaev (SYK) model \cite{Sachdev:1992fk,Kitaev:2015talk}, which constitutes a paradigmatic example of a strongly interacting quantum many-body system exhibiting quantum chaos. The model consists of $N$ Majorana fermions interacting through random, all-to-all $p$-body couplings. The SYK model provides a solvable setting for studying operator growth, Krylov dynamics, thermalization, and their symmetry-resolved structure. The Hamiltonian of the Majorana SYK model is given by
\begin{equation}\label{eq:SYK_Hamiltonian}
    H = i^{p/2} \sum_{1 \leq i_1 < i_2 < \cdots < i_p \leq N}
    J_{i_1 i_2 \cdots i_p} \, \psi_{i_1} \psi_{i_2} \cdots \psi_{i_p} \ ,
\end{equation}
where $\psi_i$ are Hermitian Majorana fermions satisfying the Clifford algebra
\begin{equation}
    \{\psi_i,\psi_j\} = \delta_{ij} \, , \qquad i,j = 1,2,\ldots,N \ ,
\end{equation}
and $p$ is an even integer; for the numerical analysis, we restrict ourselves to the quartic model, $p=4$, and take the number of Majorana fermions $N$ to be even. The phase
$i^{p/2}$ in \eqref{eq:SYK_Hamiltonian} ensures that the Hamiltonian is Hermitian for any even value of $p$. For even $N$, the Clifford algebra generated by the Majorana operators admits an irreducible representation on a Hilbert space of dimension
\begin{equation}
    D=\dim\mathcal{H}=2^{N/2} \ .
    \label{eq:real-syk-hilbert-dimension}
\end{equation}
This representation may be constructed by pairing the $N$ Majorana fermions into $N/2$ complex fermionic modes, or, equivalently, by employing a Jordan--Wigner representation in terms of $N/2$ two-level systems.

The random couplings $J_{i_1 i_2 \cdots i_p}$ are drawn from an independent real Gaussian distribution with zero mean
\begin{equation}
    \big\langle J_{i_1i_2\cdots i_p}\big\rangle
    =0 \ ,
\end{equation}
and variance
\begin{equation}
    \big\langle
    J_{i_1i_2\cdots i_p}^{\,2}
    \big\rangle
    =
    \frac{(p-1)!}{N^{p-1}}\,J^2 \ .
    \label{eq:SYK_disorder_variance}
\end{equation}
Equivalently, the couplings may be regarded as components of a completely antisymmetric rank-$p$ tensor; the restriction $i_1<i_2<\cdots<i_p$ then ensures that each independent
interaction term is included exactly once. Here, $\langle\cdots\rangle$ denotes an average over
independent realizations of the random couplings, while $J$ sets the characteristic energy scale of the model. 

Specializing to the quartic case, $p=4$, the Hamiltonian becomes
\begin{equation}
    H_{\mathrm{SYK}_4}
    =
    -\sum_{1\leq i<j<k<l\leq N}
    J_{ijkl}\,
    \psi_i\psi_j\psi_k\psi_l \ ,
    \label{eq:SYK4_Hamiltonian}
\end{equation}
Each interaction term therefore contains four distinct Majorana fermions selected from the complete set
$\{\psi_1,\psi_2,\ldots,\psi_N\}$. 

The quartic Hamiltonian possesses an exact $\mathbb{Z}_2$ fermion-parity symmetry. The corresponding parity operator is
\begin{equation}\label{eq:SYK_parity}
    \Gamma = (2i)^{N/2} \prod_{i=1}^N \psi_i \ ,
\end{equation}
It satisfies
\begin{equation}
    \Gamma^\dagger=\Gamma \ ,
    \qquad
    \Gamma^2=\mathbbm{1} \ ,
    \qquad
    \Gamma\psi_i\Gamma=-\psi_i \ .
    \label{eq:real-syk-parity-properties}
\end{equation}
Thus, fermion parity reverses the sign of every individual Majorana operator. Since each term in the SYK$_4$ Hamiltonian contains four Majorana operators, the four minus signs cancel, and hence
\begin{equation}
    [H,\Gamma]=0 \ .
    \label{eq:real-syk-parity-conservation}
\end{equation}
The Hamiltonian can therefore be block diagonalized into even and odd fermion parity sectors. Introducing the projection operators
\begin{equation}
    \Pi_\pm=\frac{1}{2}\left(\mathbbm{1}\pm\Gamma\right) \ ,
    \label{eq:real-syk-parity-projectors}
\end{equation}
the Hilbert space and Hamiltonian decompose as
\begin{equation}
    \mathcal{H}
    =
    \mathcal{H}_+\oplus\mathcal{H}_- \ ,
    \qquad
    H=H_+\oplus H_- \ ,
    \qquad
    H_\pm=\Pi_\pm H\Pi_\pm \ .
    \label{eq:real-syk-block-decomposition}
\end{equation}
\begin{figure}[t!]
\centering
\includegraphics[width=0.5\linewidth]{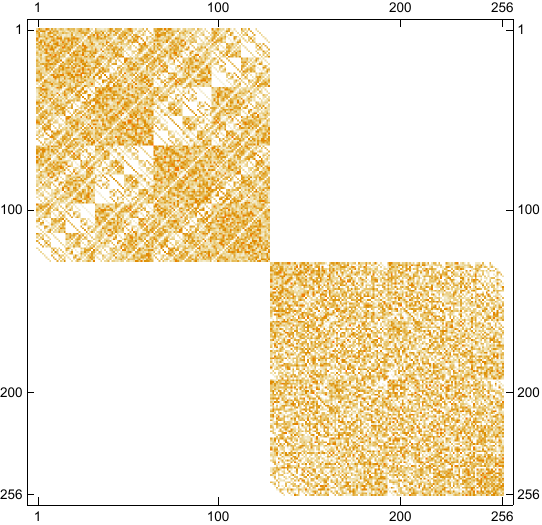}
  \par\medskip
    \includegraphics[width=0.49\linewidth]{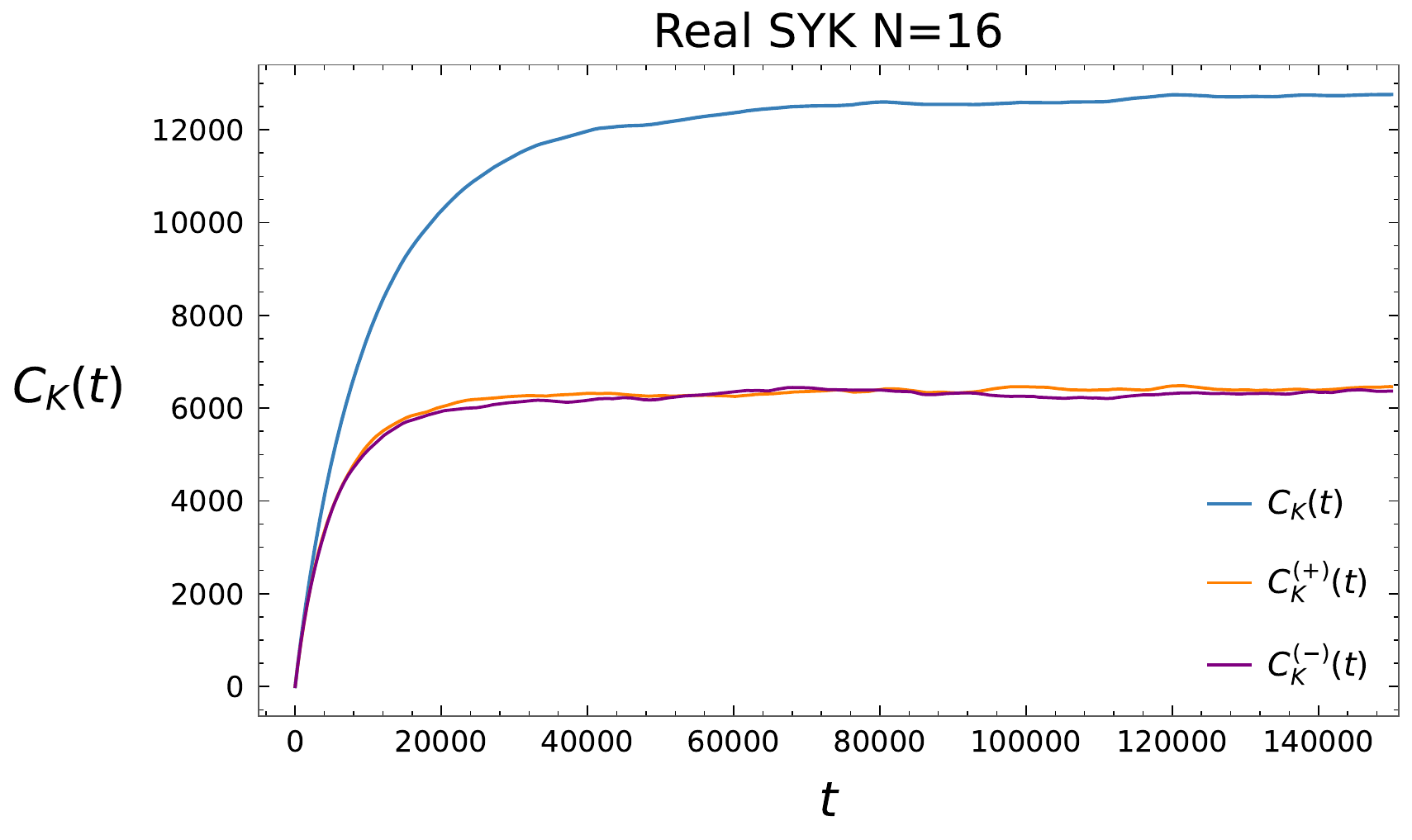}
    \hfill
    \includegraphics[width=0.49\linewidth]{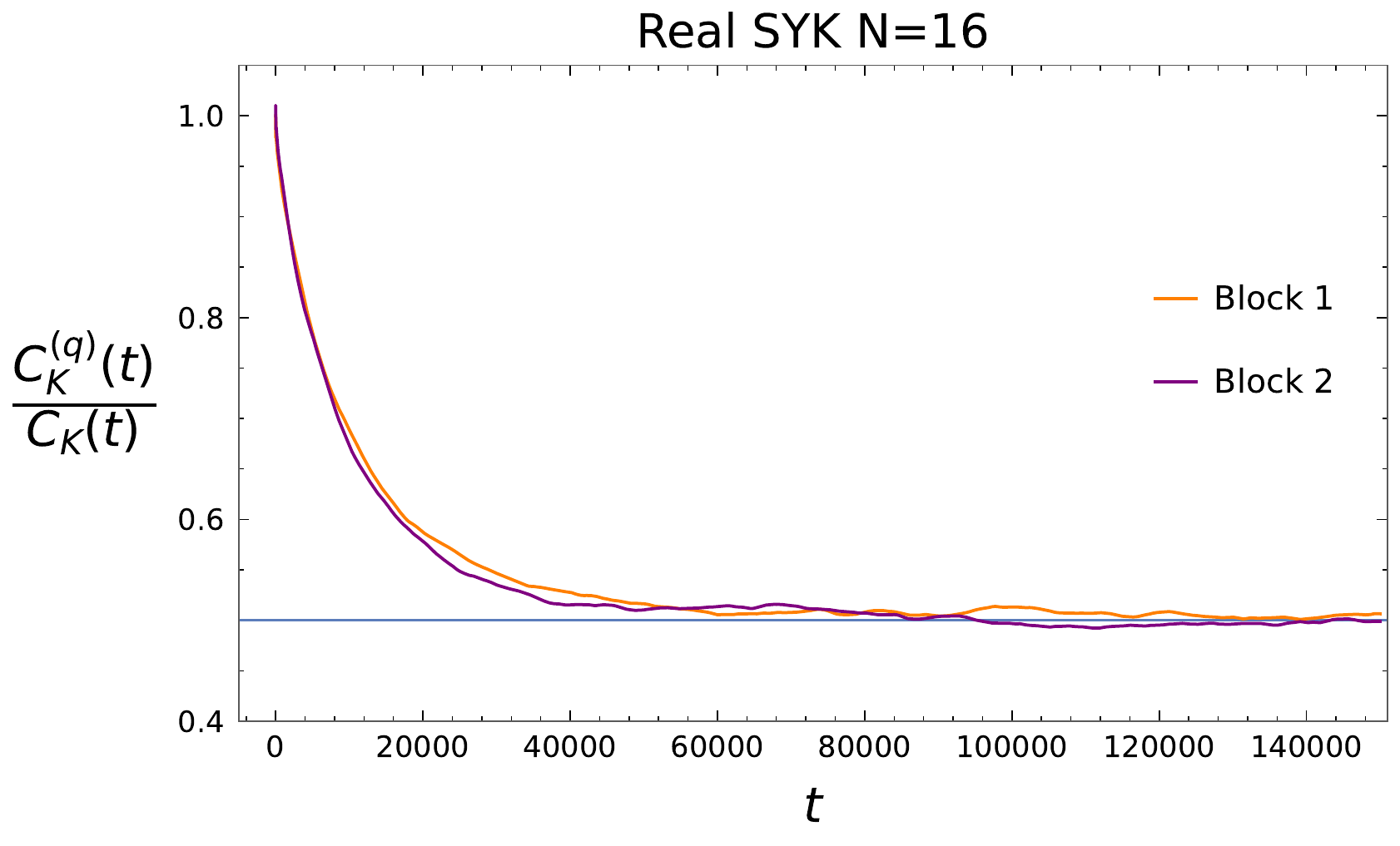}
     
    \caption{For the real SYK model with $N=16$, the upper panel shows the block-diagonal structure of the Hamiltonian in \eqref{eq:SYK4_Hamiltonian} in the fermion-parity basis. The two parity sectors have equal Hilbert-space dimensions, $d_{+}=d_{-}=2^{N/2-1}=128$. The bottom-left panel shows the time evolution of the full Krylov complexity $C_K(t)$ and the parity-resolved complexities $C_K^{(q,+)}(t)$ and $C_K^{(q,-)}(t)$. The bottom-right panel displays the corresponding fractions $C_K^{(q,\pm)}(t)/C_K(t)$.  The horizontal lines denote the expected late-time values $d_q^2/\big(\sum_q d_q^2\big)=1/2$ for both sectors, in good agreement with the numerical results.  The Krylov complexity curves are averaged over two disorder realizations.}
    \label{fig:krylov-real-syk}
\end{figure}
The two parity sectors have equal dimensions,
\begin{equation}
    d_+=d_-=2^{N/2-1} \  .
    \label{eq:real-syk-sector-dimensions}
\end{equation}
To initialize the Krylov construction, we choose the Hermitian Majorana bilinear
\begin{equation}
    O=i\psi_2\psi_3 \ ,
    \label{eq:real-syk-seed}
\end{equation}
Since
$\psi_2$ and $\psi_3$ anticommute, this operator is Hermitian.
Furthermore, the seed operator contains an even number of Majorana operators and is therefore also invariant under fermion parity,
\begin{equation}
    [O,\Gamma]=0 \ .
    \label{eq:real-syk-seed-parity}
\end{equation}
It consequently admits the same block decomposition as the Hamiltonian,
\begin{equation}
    O=O_+\oplus O_- \ ,
    \qquad
    O_\pm=\Pi_\pm O\Pi_\pm \ .
    \label{eq:real-syk-seed-blocks}
\end{equation}
Therefore the Hamiltonian and operator can be block-diagonalized simultaneously into two sectors of equal Hilbert-space dimension $2^{N/2-1}$. The Krylov construction can then be performed separately in the two parity sectors. This exact block decomposition is central to the symmetry-resolved
analysis performed below. In particular, it allows us to compare the Krylov dynamics obtained in the full operator space with that obtained after projecting onto definite fermion-parity sectors.

In Fig.~\ref{fig:krylov-real-syk} we present the block-diagonal form of the Hamiltonian \eqref{eq:SYK_Hamiltonian} for $N=16$ Majorana fermions, together with the fractional Krylov complexity $C_K^{(q)}(t)/C_K(t)$ for the two symmetry sectors. We also plot the full Krylov complexity $C_K(t)$ together with the sector-resolved complexities $C_K^{(+)}(t)$ and $C_K^{(-)}(t)$, showing their complete time evolution and late-time saturation. It is evident from the figure that, at late times, the Krylov complexity is equally distributed between the two fermion-parity sectors. More generally, this result supports the prediction that the asymptotic contribution of each symmetry sector is governed by the dimension of its
operator space, $d_q^2$, rather than by the corresponding Hilbert-space dimension, $d_q$. The numerical results therefore provide clear evidence for dimension-weighted equipartition of Krylov complexity among the symmetry sectors. In the present case, the two parity sectors have equal dimensions, $d_+=d_-$, and consequently each sector contributes one half of the total late-time Krylov complexity.

%

\subsection{Complex SYK model}
In this section, we consider the complex SYK$_4$ model, namely the quartic Sachdev--Ye--Kitaev model constructed from complex fermions \cite{Rabinovici:2020ryf}. As in the Majorana SYK model, the interactions are random and all-to-all. However, the complex model possesses an additional global $U(1)$ symmetry associated with the conservation of
the total fermion number. It therefore provides a natural setting for studying charge-resolved Krylov dynamics. The Hamiltonian is given by
\begin{equation}\label{eq:ham_complexSYK}
H=
\sum_{ijkl}
J_{ij;kl} \ 
c_i^\dagger c_j^\dagger c_k c_l
+\text{h.c.} \ ,
\end{equation}
where $i,j,k,l=1,2 \cdots N$ and h.c. denotes the Hermitian conjugate and ensures that $H=H^\dagger$. The operators $c_i$ and $c_i^\dagger$ are complex fermion annihilation and creation operators
satisfying the canonical anticommutation relations
\begin{equation}
\{c_i,c_j^\dagger\}=\delta_{ij} \ ,
\qquad
\{c_i,c_j\}=\{c_i^\dagger,c_j^\dagger\}=0 \ ,
\qquad i,j=1,2,\ldots,N \ .
\end{equation}
Each of the $N$ complex-fermion modes can be either empty or occupied. The corresponding Fock-space basis is therefore
\begin{equation}
    \ket{n_1,n_2,\ldots,n_N} \ ,
    \qquad
    n_i\in\{0,1\} \ ,
\end{equation}
The Hilbert-space dimension is therefore $D=2^N$.\footnote{Notice that, in contrast to the real SYK model, where $N$ Majorana fermions generate a Hilbert space of dimension $2^{N/2}$ for even $N$, here $N$ counts the number of complex-fermion modes themselves and hence gives a Hilbert-space dimension $2^N$.} The random couplings $J_{ij;kl}$ are independent complex Gaussian variables with zero mean. In our numerical implementation, we take
\begin{equation}
    J^{(R)}_{ij;kl},\,J^{(I)}_{ij;kl}
    \sim
    \mathcal{N}\left(
        0,\frac{3!}{2}\frac{J^2}{N^3}
    \right) \ .
    \label{eq:complexSYK_gaussian_distribution}
\end{equation}
Consequently, the disorder averages satisfy
\begin{equation}
    \big\langle J_{ij;kl}\big\rangle=0 \ ,
    \qquad
    \big\langle |J_{ij;kl}|^2\big\rangle
    =
    \frac{3!J^2}{N^3} \ .
    \label{eq:complexSYK_disorder_variance}
\end{equation}
In addition, we set $J_{ij;ij}=0$, thereby excluding diagonal density-density terms. Here, $\langle\cdots\rangle$ denotes the average over independent disorder realizations, while $J$ sets the characteristic interaction energy scale.

The complex SYK$_4$ Hamiltonian possesses an exact global $U(1)$ symmetry generated by the total fermion-number operator
\begin{equation}
    Q=\sum_{i=1}^{N}c_i^\dagger c_i \ .
    \label{eq:complex-syk-charge}
\end{equation}
The finite transformation generated by $Q$ is
\begin{equation}
    U(\alpha)=e^{i\alpha Q} \ .
\end{equation}
Under this transformation, each fermionic creation operator acquires a phase, while each annihilation operator acquires the opposite phase. Every quartic interaction term in the Hamiltonian contains two creation and two annihilation operators. The phases therefore cancel exactly, leaving each interaction term unchanged. The same argument applies to its Hermitian conjugate. Physically, the interaction only scatters pairs of fermions between different modes and does not change the total fermion number. Consequently, the complete Hamiltonian is invariant under the global $U(1)$ transformation,
\begin{equation}
    U(\alpha)HU^\dagger(\alpha)=H \ ,
\end{equation}
or, equivalently,
\begin{equation}
    [H,Q]=0 \ .
    \label{eq:complex-syk-charge-conservation}
\end{equation}
Therefore, the total fermion number is conserved under time evolution. This implies that the Hamiltonian cannot connect states carrying different eigenvalues of $Q$. Indeed, if $\ket{\phi_q}$ is an eigenstate of $Q$ with eigenvalue $q$, then
\begin{equation}
    Q\ket{\phi_q}=q\ket{\phi_q} \ ,
\end{equation}
and, using $[H,Q]=0$, one finds
\begin{align}
    QH\ket{\phi_q}
    &=
    HQ\ket{\phi_q}
    = qH\ket{\phi_q} \ .
    \label{eq:complex-syk-sector-preservation}
\end{align}
Therefore, $H\ket{\phi_q}$ remains in the same charge-$q$ sector.

To make this decomposition explicit, consider the occupation-number basis
\begin{equation}
    \ket{\boldsymbol{n}}
    =
    \ket{n_1,n_2,\ldots,n_N} \ ,
    \qquad
    n_i\in\{0,1\} \ .
\end{equation}
The total fermion-number operator acts on these states according to
\begin{equation}
    Q\ket{n_1,n_2,\ldots,n_N}
    =
    \left(\sum_{i=1}^{N}n_i\right)
    \ket{n_1,n_2,\ldots,n_N} \ .
    \label{eq:complex-syk-charge-eigenvalue}
\end{equation}
Since each occupation number is either zero or one, the possible eigenvalues of $Q$ are
\begin{equation}
    q=0,1,\ldots,N \ .
\end{equation}
The full Hilbert space consequently decomposes into $N+1$ eigenspaces of definite fermion number,
\begin{equation}
    \mathcal{H}
    =
    \bigoplus_{q=0}^{N}\mathcal{H}_q \ ,
    \label{eq:complex-syk-hilbert-decomposition}
\end{equation}
where
\begin{equation}
    \mathcal{H}_q
    =
    \operatorname{span}
    \left\{
        \ket{n_1,n_2,\ldots,n_N}
        \,\middle|\,
        \sum_{i=1}^{N}n_i=q
    \right\} \ .
    \label{eq:complex-syk-charge-sector}
\end{equation}
A basis state in $\mathcal{H}_q$ is obtained by choosing exactly $q$ occupied modes from the available $N$ complex-fermion modes. The number of such choices is therefore
\begin{equation}
    d_q
    =
    \dim\mathcal{H}_q
    =
    \binom{N}{q} \ .
    \label{eq:complex-syk-sector-dimension}
\end{equation}
The dimensions of all fixed-charge sectors correctly reproduce the dimension of the complete Fock space:
\begin{equation}
    \sum_{q=0}^{N}d_q
    =
    \sum_{q=0}^{N}\binom{N}{q}
    =
    2^N
    =
    \dim\mathcal{H} \ .
    \label{eq:complex-syk-sector-dimension-sum}
\end{equation}
The projection operator onto the charge-$q$ sector may be written as
\begin{equation}
    \Pi_q
    =
    \frac{1}{2\pi}
    \int_{0}^{2\pi}
    d\alpha\,
    e^{i\alpha(Q-q)} \ .
    \label{eq:complex-syk-charge-projector}
\end{equation}
These projection operators satisfy
\begin{equation}
    \Pi_q\Pi_{q'}
    =
    \delta_{qq'}\Pi_q \ ,
    \qquad
    \sum_{q=0}^{N}\Pi_q
    =
    \mathbbm{1} \ ,
    \qquad
    Q\Pi_q=q\Pi_q \ .
    \label{eq:complex-syk-projector-properties}
\end{equation}
Since the Hamiltonian commutes with $Q$, it also commutes with every $\Pi_q$. It therefore assumes the block-diagonal form
\begin{equation}
    H
    =
    \bigoplus_{q=0}^{N}H_q \ ,
    \qquad
    H_q
    =
    \Pi_qH\Pi_q \ ,
    \label{eq:complex-syk-hamiltonian-blocks}
\end{equation}
where the $q$-th Hamiltonian block acts entirely within the
$d_q$-dimensional Hilbert space $\mathcal{H}_q$.

To initialize the Krylov construction, we choose the
Hermitian, number-conserving hopping operator
\begin{equation}
    O=c_1^\dagger c_2+c_2^\dagger c_1 \ .
    \label{eq:complex-syk-seed}
\end{equation}
The seed operator is also invariant under the global $U(1)$ symmetry. 
\begin{equation}
    [Q,O]=0 \ .
    \label{eq:complex-syk-seed-charge-conservation}
\end{equation}
Equivalently, under the finite $U(1)$ transformation generated by $Q$,
\begin{equation}
    U(\alpha)OU^\dagger(\alpha)=O \ ,
    \qquad
    U(\alpha)=e^{i\alpha Q} \ .
\end{equation}
Thus, the hopping process transfers a fermion between the first and second modes without changing the total fermion number. The seed operator therefore admits the same fixed-charge decomposition as the Hamiltonian,
\begin{equation}
    O
    =
    \bigoplus_{q=0}^{N}O_q \ ,
    \qquad
    O_q=\Pi_qO\Pi_q \ .
    \label{eq:complex-syk-seed-blocks}
\end{equation}
The Krylov construction can therefore be carried out independently within each fixed-charge sector. This exact decomposition allows us to compare the Krylov dynamics in the full charge-preserving operator space with the dynamics resolved into sectors of definite fermion number. In contrast to the real SYK model, the charge sectors generally have unequal dimensions. The complex SYK model therefore provides a nontrivial test of the predicted $d_q^2$-weighted distribution of late-time Krylov complexity.
\begin{figure}[!p]
    \centering
    \vspace{2mm}

  \hspace{5mm}  \includegraphics[width=0.45\linewidth]{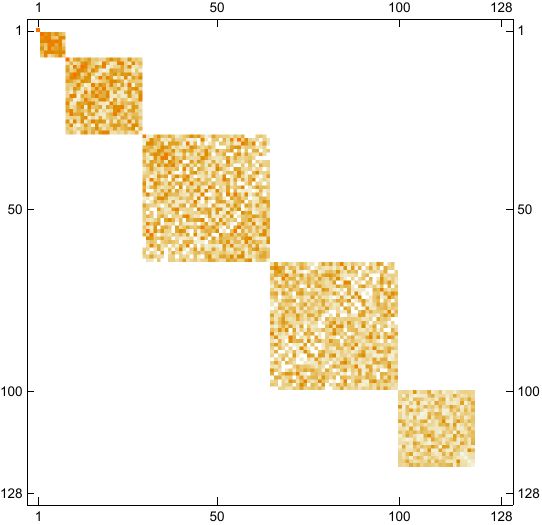}
    \hfill
    \hspace{6mm} \includegraphics[width=0.45\linewidth]{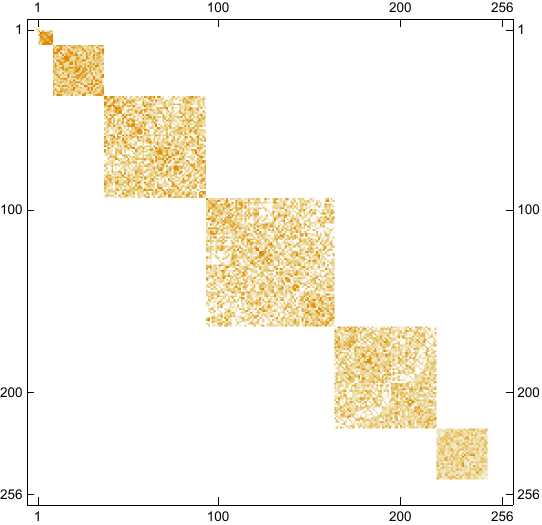}
    \\[2mm]

    \includegraphics[width=0.49\linewidth]{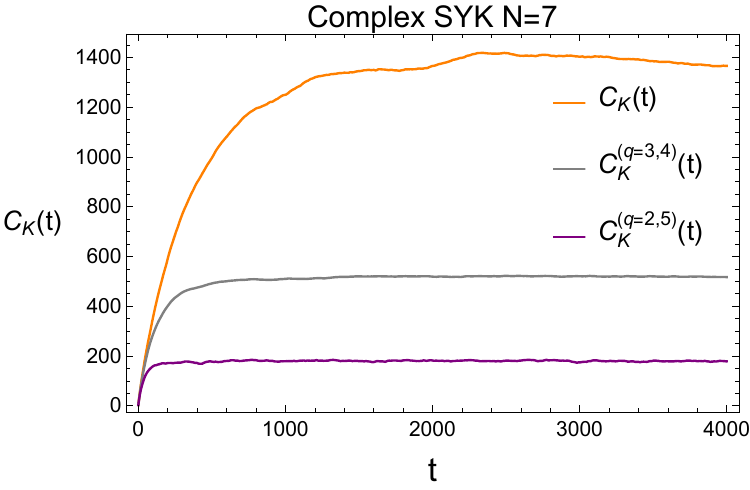}
    \hfill
    \includegraphics[width=0.49\linewidth]{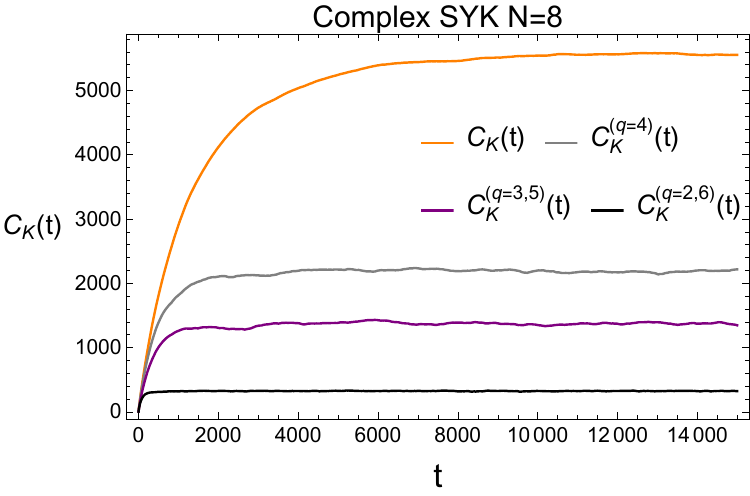}
    \\[2mm]

    \includegraphics[width=0.49\linewidth]{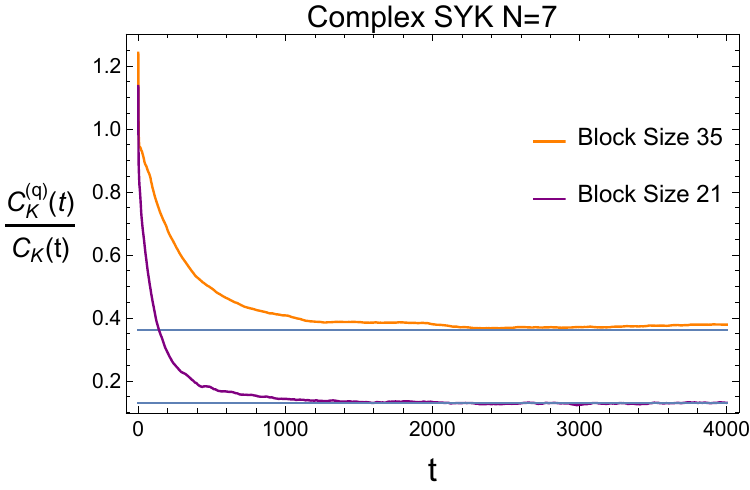}
    \hfill
    \includegraphics[width=0.49\linewidth]{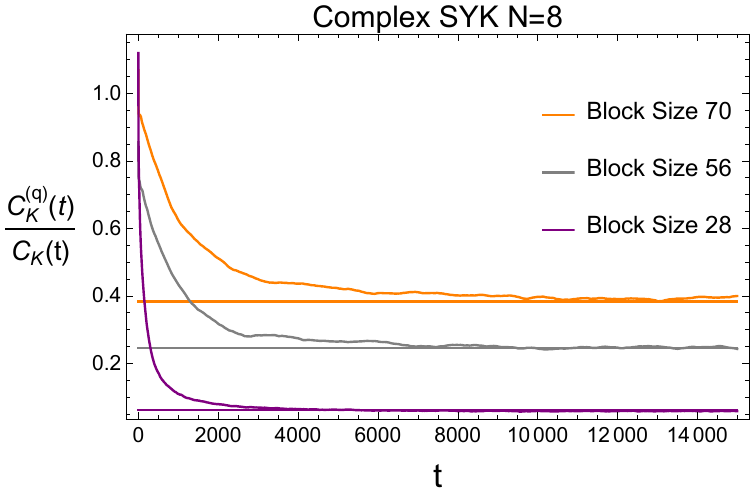}

   \caption{The upper-left and upper-right panels show the block-diagonal structure of the complex SYK Hamiltonian in \eqref{eq:ham_complexSYK}, expressed in the number-operator basis, for $N=7$ and $N=8$, respectively. The corresponding sector dimensions are $d_q=\binom{7}{q}=1,7,21,35,35,21,7,1$ for $N=7$ and $d_q=\binom{8}{q}=1,8,28,56,70,56,28,8,1$ for $N=8$. The central panels display the full Krylov complexity $C_K(t)$ together with the symmetry-resolved complexities $C_K^{(q)}(t)$ for the indicated charge sectors. The lower panels show the corresponding fractional Krylov complexities $C_K^{(q)}(t)/C_K(t)$ for block sizes $21$ and $35$ at $N=7$, and $28$, $56$, and $70$ at $N=8$. The horizontal lines indicate the expected late-time values $d_q^2/\bigl(\sum_{q'}d_{q'}^2\bigr)$, which are in good agreement with the numerical results. For \(N=7\) and \(N=8\), the Krylov complexity curves are averaged over \(50\) and \(10\) disorder realizations respectively.}
    \label{fig:krylov-complex-syk}
\end{figure}

In Fig.~\ref{fig:krylov-complex-syk}, we present the block-diagonal structure of the complex SYK$_4$ Hamiltonian \eqref{eq:ham_complexSYK} for $N=7$ and $N=8$ complex fermions, together with the corresponding fractional Krylov complexities $C_K^{(q)}(t)/C_K(t)$ in the fixed-charge sectors. We also plot the full Krylov complexity $C_K(t)$ together with the sector-resolved complexities $C_K^{(q)}(t)$, showing their complete time evolution and late-time saturation; particle--hole-related sectors $q$ and $N-q$ exhibit identical dynamics. At late times, the Krylov complexity is distributed among the charge sectors according to their operator-space dimensions. More precisely, the asymptotic contribution of the charge-$q$ sector is governed by $d_q^2=\binom{N}{q}^{2}$ rather than by its Hilbert-space dimension $d_q=\binom{N}{q}$ alone. Consequently, sectors with larger Hilbert-space dimensions contribute a larger fraction of the total late-time complexity. Moreover, the particle--hole-related sectors $q$ and $N-q$ have equal dimensions and therefore make equal asymptotic contributions. The numerical results for both $N=7$ and $N=8$ are in good agreement with this prediction, providing clear evidence for dimension-weighted equipartition of Krylov complexity among the fixed-charge sectors.

For the Hamiltonian and seed operator considered here, the sectors \(q=0,1,N\) require special treatment. In the completely empty sector, the quartic Hamiltonian cannot act because there are no fermions to annihilate, and hence \(H_0=0\). Moreover, since every term in the Hamiltonian contains two annihilation operators, it also annihilates
every one-particle state, so that \(H_1=0\). Although the hopping seed \(O=c_1^\dagger c_2+c_2^\dagger c_1\) is generally nonzero in the one-particle sector, the corresponding Liouvillian vanishes, \(\mathcal{L}_1=[H_1,\cdot]=0\). The seed vanishes in both extreme charge sectors: \(O_0=0\), since there is no fermion available to hop, and \(O_N=0\), since there is no unoccupied mode into which a fermion can hop. Thus, the sectors \(q=0,1,N\) produce no nontrivial Krylov dynamics. Therefore, restricting the analysis to the contributing charge sectors, the predicted late-time fraction is 
\begin{equation}
    \frac{C_K^{(q)}(t)}{C_K(t)}
    \xrightarrow[t\to\infty]{}
    \frac{d_q^2}
    {\displaystyle\sum_{q'=2}^{N-1}d_{q'}^2}
    =
    \frac{\binom{N}{q}^{2}}
    {\displaystyle\sum_{q'=2}^{N-1}\binom{N}{q'}^{2}} \ ,
    \qquad q=2,\ldots,N-1 \ .
    \label{eq:complex-syk-late-time-fraction}
\end{equation}
%

 \subsection{Bosonic spin model}\label{subsec:Bosonicmodel}

In this section, we consider a bosonic quadratic chaotic spin model, which is a simplified version of the model introduced in~\cite{Basu:2025ubf}. The  Hamiltonian is given by
 \begin{equation}\label{eq:ham_4block}
     H=\sqrt{\frac{1}{2N}} 
 \sum_{1 \leq i_1 < i_2 \leq N}
  \left(J_{i_1 i_2}^x S_{i_1}^x S_{i_2}^x +J_{i_1 i_2}^y S_{i_1}^y S_{i_2}^y \right) \ .
 \end{equation}
The couplings \(J_{i_1i_2}^{x}\) and \(J_{i_1i_2}^{y}\) are independent real Gaussian random variables with zero mean and unit variance. 

The local spin operators are defined as
 \begin{align}
 S_i^\alpha = \mathbb{I}^{\otimes (i-1)} \otimes \sigma^\alpha \otimes \mathbb{I}^{\otimes (N-i)} \ ,
 \qquad \alpha \in \{x,y,z\}, \quad i=1,2,\ldots,N \ ,
 \end{align}
 where $\mathbb{I}$ denotes the $2 \times 2$ identity matrix, $\sigma^\alpha$ are the Pauli matrices, and $N$ is the total number of spin sites. The full Hilbert space therefore has dimension
 \begin{equation}
     \dim\mathcal{H}=2^N \ .
 \end{equation}
 The model possesses global $\mathbb{Z}_2$ symmetries generated by
 \begin{equation}
     \Gamma_z
    =
    \bigotimes_{i=1}^{N}\sigma_i^z \ ,
     \qquad
    \Gamma_x
     =
    \bigotimes_{i=1}^{N}\sigma_i^x \ .
     \label{eq:spin-global-charges}
 \end{equation}
 Indeed, conjugation by $\Gamma_z$ changes the signs of both $S_i^x$ and $S_i^y$,
 \begin{equation}
     \Gamma_zS_i^x\Gamma_z=-S_i^x \ ,
     \qquad
     \Gamma_zS_i^y\Gamma_z=-S_i^y \ ,
 \end{equation}
 so that every quadratic term in the Hamiltonian remains invariant. Similarly, under conjugation by $\Gamma_x$ ,
 \begin{equation}
     \Gamma_xS_i^x\Gamma_x=S_i^x \ ,
    \qquad
     \Gamma_xS_i^y\Gamma_x=-S_i^y \ ,
 \end{equation}
 and the Hamiltonian is again invariant. Consequently,
 \begin{equation}
    [H,\Gamma_z]=0 \ ,
    \qquad
    [H,\Gamma_x]=0 \ .
     \label{eq:spin-hamiltonian-symmetries}
     \end{equation}
 Although the Hamiltonian commutes with both generators for any $N$, the two symmetry operators satisfy
 \begin{equation}
     \Gamma_z\Gamma_x
     =
    (-1)^N\Gamma_x\Gamma_z \ .
 \end{equation}
 They can therefore be simultaneously diagonalized when $N$ is even and together they form a pair of conserved charges.
 The corresponding $\Gamma_y$ symmetry is not independent, since $\Gamma_y$ is proportional to the product $\Gamma_x\Gamma_z$.

 For even $N$, the simultaneous eigenvalues of $\Gamma_x$ and
 $\Gamma_z$ are denoted by
 \begin{equation}
     s_x=\pm1 \ ,
     \qquad
     s_z=\pm1 \ .
 \end{equation}
 The projector onto the sector labeled by $(s_x,s_z)$ is
 \begin{equation}
     \Pi_{s_x,s_z}
     =
     \frac{1}{4}
    \left(\mathbb{I}+s_x\Gamma_x\right)
     \left(\mathbb{I}+s_z\Gamma_z\right) \ .
     \label{eq:spin-sector-projector}
 \end{equation}
 The Hilbert space consequently decomposes into four symmetry sectors,
 \begin{equation}
     \mathcal{H}
     =
     \bigoplus_{s_x=\pm1}
     \bigoplus_{s_z=\pm1}
     \mathcal{H}_{s_x,s_z} \ ,
     \label{eq:spin-hilbert-decomposition}
 \end{equation}
 where
 \begin{equation}
     \mathcal{H}_{s_x,s_z}
     =
     \left\{
         \ket{\psi}\in\mathcal{H}
         \,\middle|\,
         \Gamma_x\ket{\psi}=s_x\ket{\psi},
         \quad
         \Gamma_z\ket{\psi}=s_z\ket{\psi}
     \right\} \ .
 \end{equation}
 All four sectors have the same Hilbert-space dimension,
 \begin{equation}
     d_{s_x,s_z}
     =
     \dim\mathcal{H}_{s_x,s_z}
     =
     2^{N-2} \ ,
     \label{eq:spin-sector-dimension}
 \end{equation}
 and the Hamiltonian assumes the block-diagonal form
 \begin{equation}
     H
     =
     \bigoplus_{s_x=\pm1}
     \bigoplus_{s_z=\pm1}
     H_{s_x,s_z} \ ,
     \qquad
     H_{s_x,s_z}
     =
     \Pi_{s_x,s_z}H\Pi_{s_x,s_z} \ .
     \label{eq:spin-hamiltonian-blocks}
 \end{equation}
 \begin{figure}[t!]
\centering
\includegraphics[width=0.45\linewidth]{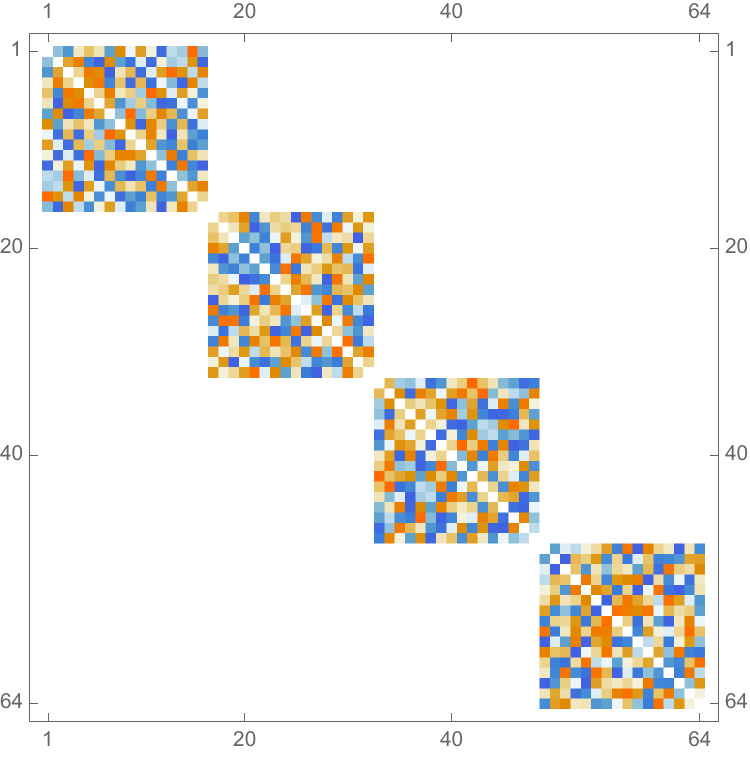}
\par\medskip
    \includegraphics[width=0.49\linewidth]{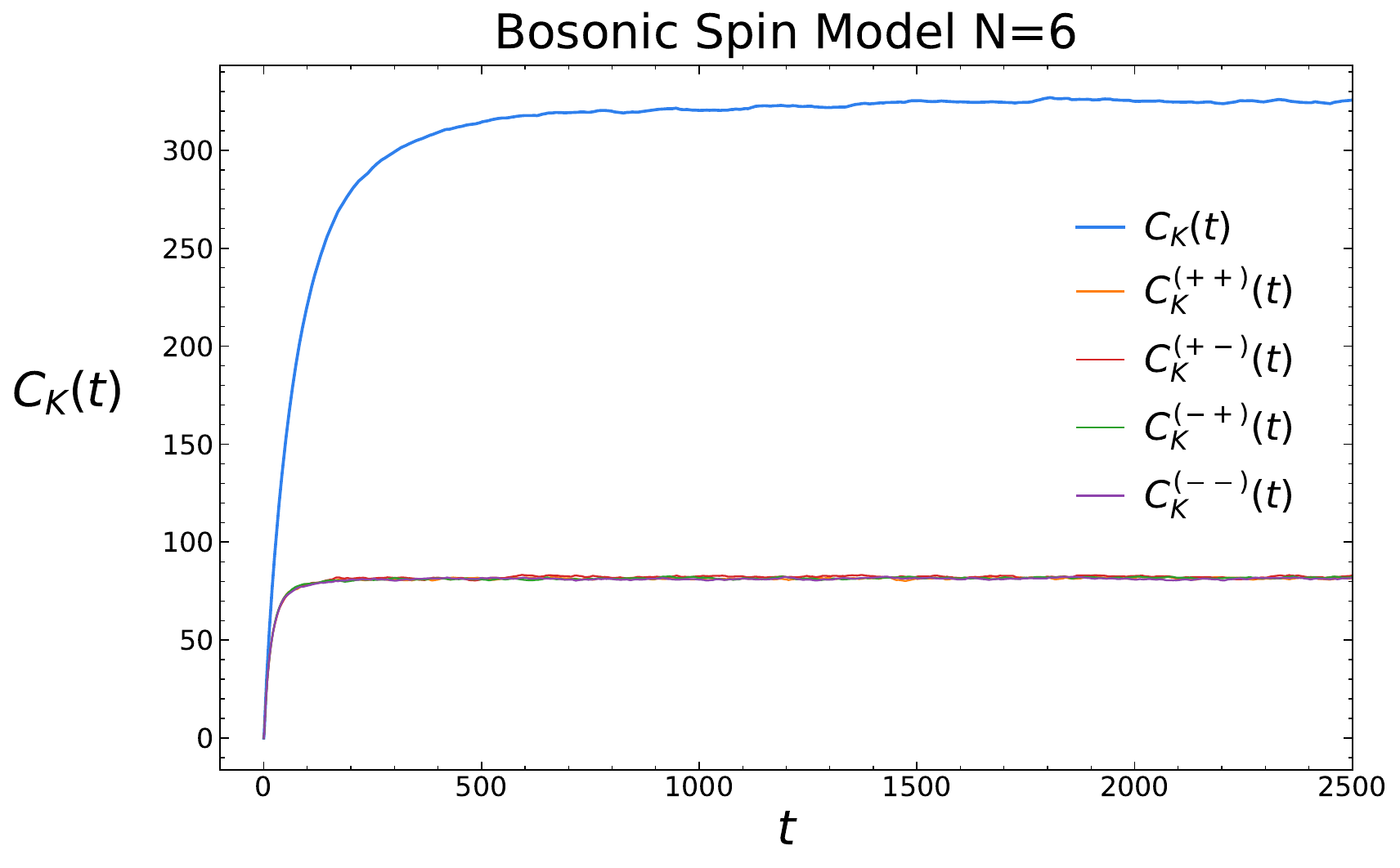}
    \hfill
    \includegraphics[width=0.49\linewidth]{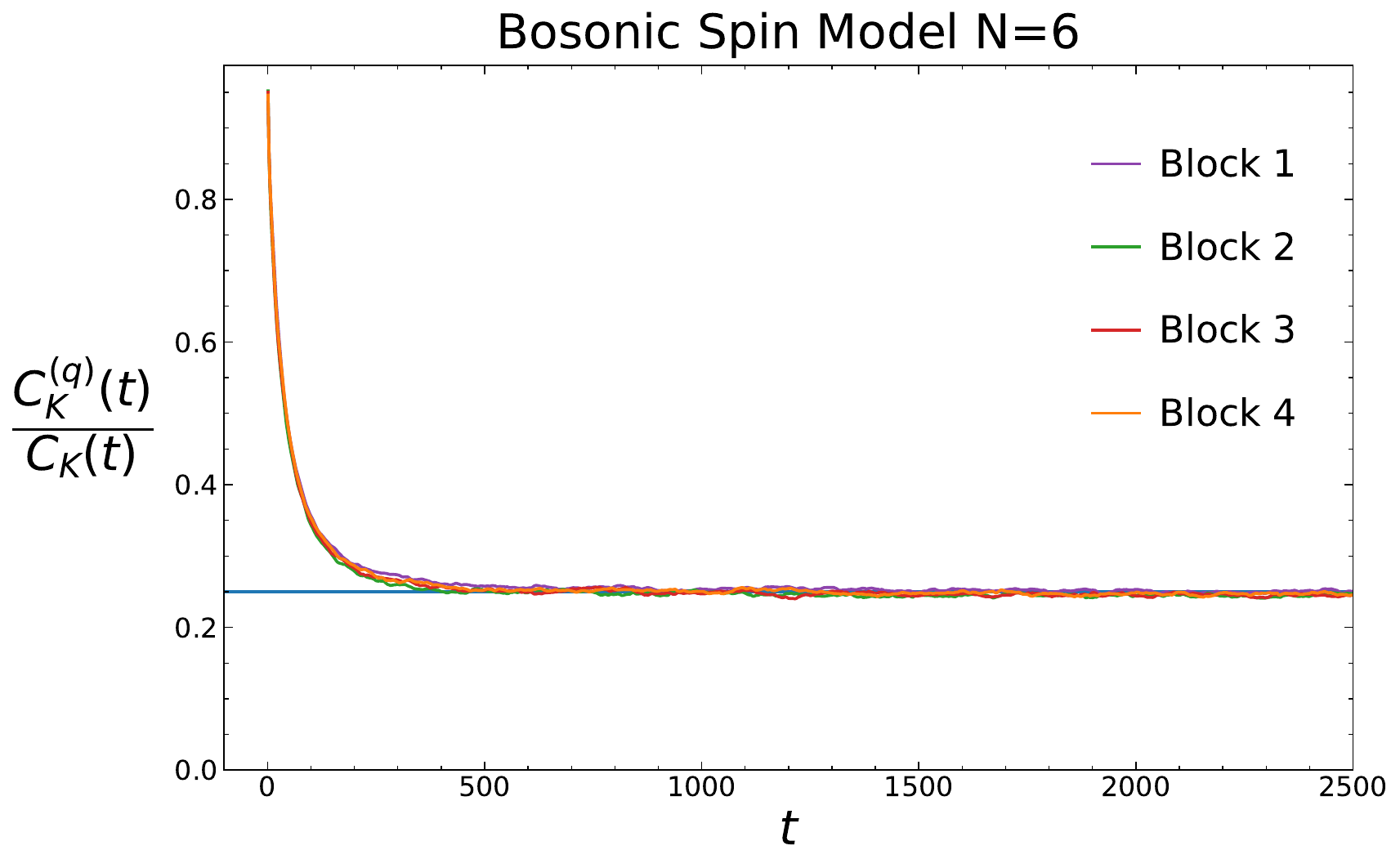}
       \caption{The upper panel shows the block-diagonal structure of the Hamiltonian in Eq.~\eqref{eq:ham_4block} in the simultaneous \((\Gamma_x,\Gamma_z)\) symmetry eigenbasis. All four symmetry sectors have equal Hilbert-space dimension \(d_q=2^{N-2}=16\). The bottom-left panel shows the time evolution of the full Krylov complexity \(C_K(t)\) and the symmetry-resolved complexities \(C_K^{(q)}(t)\) for the four \(\mathbb{Z}_2\times\mathbb{Z}_2\) sectors \(q=(++),(+-),(-+),(--)\), for the bosonic spin model with \(N=6\). The bottom-right panel displays the corresponding fractional complexities \(C_K^{(q)}(t)/C_K(t)\). The horizontal line denotes the predicted late-time value \(d_q^2/\sum_{q'}d_{q'}^2=1/4\), which agrees well with the numerical results. The Krylov complexity curves are averaged over $100$ disorder realizations. }
    \label{fig:krylov-spin}
 \end{figure} 

 To initialize the Krylov construction, we choose the Hermitian seed
 operator
 \begin{equation}
     O=S_1^xS_2^x \ .
     \label{eq:spin-seed}
 \end{equation}
 Since $O$ contains two $x$-spin operators, the two sign changes generated by $\Gamma_z$ cancel, while conjugation by $\Gamma_x$ leaves each $S_i^x$ invariant. Thus,
 \begin{equation}
     [O,\Gamma_z]=0 \ ,
     \qquad
     [O,\Gamma_x]=0 \ .
     \label{eq:spin-seed-symmetries}
 \end{equation}
 The seed operator therefore preserves each symmetry sector and admits the same block decomposition as the Hamiltonian,
 \begin{equation}
     O
     =
     \bigoplus_{s_x=\pm1}
     \bigoplus_{s_z=\pm1}
     O_{s_x,s_z} \ ,
     \qquad
     O_{s_x,s_z}
     =
     \Pi_{s_x,s_z}O\Pi_{s_x,s_z} \ .
     \label{eq:spin-seed-blocks}
 \end{equation}
In Fig.~\ref{fig:krylov-spin}, we present the block-diagonal form of the Hamiltonian~\eqref{eq:ham_4block} for the bosonic spin model with \(N=6\) sites, together with the fractional Krylov complexities \(C_K^{(q)}(t)/C_K(t)\) for the four \(\mathbb{Z}_2\times\mathbb{Z}_2\) symmetry sectors. We also plot the full Krylov complexity \(C_K(t)\), together with the sector-resolved complexities \(C_K^{(++)}(t)\), \(C_K^{(+-)}(t)\), \(C_K^{(-+)}(t)\), and \(C_K^{(--)}(t)\), showing their complete time evolution and late-time saturation. It is evident from the figure that, at late times, the Krylov complexity is equally distributed among the four symmetry sectors. The asymptotic contribution of each sector is governed by the dimension of its operator space, \(d_q^2\), rather than by the corresponding Hilbert-space dimension \(d_q\). For \(N=6\), all four symmetry sectors have the same Hilbert-space dimension. Consequently, the predicted late-time fractions are\(\frac{d_q^2}{\sum_{q'}d_{q'}^2} = \frac{1}{4}.\) The numerical results converge to these predicted values at late times, thereby confirming that the Krylov complexity is distributed among the symmetry sectors according to their operator-space dimensions.

\subsection{The mixed-field Ising model}\label{subsec:Mixedfield}

In this section, we consider the mixed-field
Ising chain with open boundary conditions. Its Hamiltonian is given by \cite{Camargo:2024deu,Bhattacharya:2023xjx}
\begin{equation}
H = - \sum_{i=1}^{N-1} S_i^z S_{i+1}^z 
- \sum_{i=1}^{N} \left( h_x S_i^x + h_z S_i^z \right) \ ,
\label{eq:MFIM_Hamiltonian}
\end{equation}
where \(N\) denotes the number of lattice sites, while \(h_x\) and \(h_z\) are the transverse and longitudinal magnetic fields, respectively. The local Pauli operators are embedded in the full many-body Hilbert space according to
\begin{equation}
    S_i^k
    =
    \mathbb{I}_2^{\otimes(i-1)}
    \otimes \sigma^k
    \otimes
    \mathbb{I}_2^{\otimes(N-i)} \ ,
    \qquad
    k\in\{x,y,z\} \ ,
    \label{eq:spin_operators}
\end{equation}
where \(\sigma^k\) are the Pauli matrices. The Hilbert space of the spin chain is 
\begin{equation}
    \mathcal{H}
    =
    \bigotimes_{i=1}^{N}\mathbb{C}^2 \ ,
    \qquad
    D\equiv\dim\mathcal{H}=2^N \ .
\end{equation}
The model becomes integrable when either the transverse or the
longitudinal field vanishes. In the following, we instead work at
\begin{equation}
    (h_x,h_z)=(-1.05,\,0.5) \ ,
    \label{eq:MFIM_parameters}
\end{equation}
which lies in the non-integrable, quantum-chaotic regime.

Although the simultaneous presence of transverse and longitudinal fields breaks the usual spin-flip symmetry, the Hamiltonian remains invariant under reflection about the center of the chain. The corresponding reflection operator reverses the ordering of the lattice sites,
\begin{equation}
    \Pi\,
    \ket{s_1,s_2,\ldots,s_N}
    =
    \ket{s_N,s_{N-1},\ldots,s_1} \ ,
\end{equation}
and may be written explicitly as a product of pairwise permutation
operators:
\begin{equation}
    \Pi
    =
    \begin{cases}
    \displaystyle
    \hat{P}_{1,N}\hat{P}_{2,N-1}\cdots
    \hat{P}_{\frac{N}{2},\,\frac{N}{2}+1} \ ,
    & N\ \text{even} \ ,
    \\[8pt]
    \displaystyle
    \hat{P}_{1,N}\hat{P}_{2,N-1}\cdots
    \hat{P}_{\frac{N-1}{2},\,\frac{N+3}{2}} \ ,
    & N\ \text{odd} \ .
    \end{cases}
    \label{eq:parity_operator}
\end{equation}
In the normalization adopted in Eq.~\eqref{eq:spin_operators}, the permutation operator exchanging the states at sites \(i\) and \(j\) takes the form
\begin{equation}
\hat{P}_{i,j} = \frac{1}{2} \left( \mathbb{I} + S_i^x S_j^x + S_i^y S_j^y + S_i^z S_j^z \right) \ .
\label{eq:permutation_operator}
\end{equation}
The reflection operator is Hermitian and involutive,
\begin{equation}
    \Pi^\dagger=\Pi \ ,
    \qquad
    \Pi^2=\mathbb{I}\ ,
\end{equation}
and therefore has eigenvalues \(q=\pm1\). Moreover,
\begin{equation}
    [H,\Pi]=0 \ ,
\end{equation}
so that the Hilbert space decomposes into reflection-even and
reflection-odd sectors,
\begin{equation}
    \mathcal{H}
    =
    \mathcal{H}_{+}\oplus\mathcal{H}_{-} \ ,
    \qquad
    H=H_{+}\oplus H_{-} \ .
    \label{eq:MFIM_sector_decomposition}
\end{equation}
The corresponding projection operators are
\begin{equation}
    \mathcal{P}_{q}
    =
    \frac{1}{2}
    \left(
        \mathbbm{1}+q \, \Pi
    \right),
    \qquad
    q=\pm1 \ ,
    \label{eq:MFIM_parity_projectors}
\end{equation}
with
\begin{equation}
    H_q=\mathcal{P}_qH\mathcal{P}_q \ .
\end{equation}

The dimensions \(d_q=\dim\mathcal{H}_q\) of the two parity sectors
follow from
\begin{equation}
    d_q
    =
    \rm Tr\mathcal{P}_q
    =
    \frac{1}{2}
    \left(
        2^N+q\,\rm Tr \, \Pi
    \right) \ .
\end{equation}
The trace of the reflection operator may be evaluated in the computational basis,
\begin{equation}
    \rm Tr\Pi
    =
    \sum_{\{s_i\}}
    \bra{s_1,\ldots,s_N}
    \Pi
    \ket{s_1,\ldots,s_N} \ .
\end{equation}
Since reflection reverses the ordering of the spins, a basis state contributes to $\rm Tr\Pi$ only if it is invariant under
reflection, namely if
\begin{equation}
    s_i=s_{N+1-i} \ ,
    \qquad i=1,\ldots,N \ .
\end{equation}
For an even chain, $N=2m$, the spins on the first $m$ sites can be chosen independently, while reflection uniquely fixes the remaining $m$ spins. There are therefore $2^m=2^{N/2}$ reflection-invariant configurations. For an odd chain, $N=2m+1$, the first $m$ spins and the central spin can be chosen independently, while the other $m$ spins are fixed by reflection. In this case, the number of invariant
configurations is $2^{m+1}=2^{(N+1)/2}$. These two results can be written compactly as
\begin{equation}
    \rm Tr\Pi
    =
    \begin{cases}
        2^{N/2}, & N\ \text{even},\\[2pt]
        2^{(N+1)/2}, & N\ \text{odd},
    \end{cases}
    =
    2^{\lceil N/2\rceil} \ ,
\end{equation}
where $\lceil N/2\rceil$ denotes the smallest integer greater than or equal to $N/2$. Consequently,
\begin{equation}
    d_{\pm}
    =
    \rm Tr\mathcal{P}_{\pm}
    =
    \frac{1}{2}
    \left(
        2^N\pm2^{\lceil N/2\rceil}
    \right)
    =
    2^{N-1}
    \pm
    2^{\lceil N/2\rceil-1} \ .
    \label{eq:MFIM_sector_dimensions}
\end{equation}
The inequality $d_{+}>d_{-}$ arises because every configuration that is not reflection invariant can be combined with its reflected partner to produce one even and one odd state, whereas a reflection-invariant configuration contributes only to the even sector.

Unlike the fermion-parity sectors of the real SYK model, the two reflection sectors of the mixed-field Ising chain therefore have unequal dimensions. This makes the model particularly useful for testing whether the late-time Krylov complexity is equally shared between symmetry sectors or is instead controlled by the dimensions of their accessible operator spaces.

In the remainder of this section, we restrict to chains with an even number of sites, \(N=2m\), and perform the numerical analysis for
\(N=8\).

For the \(N=8\) chain considered numerically below, one finds
\begin{equation}
    \operatorname{Tr}\Pi=2^4=16 \ ,
    \qquad
    d_{+}=136 \ ,
    \qquad
    d_{-}=120 \ .
    \label{eq:MFIM_N8_dimensions}
\end{equation}
For an even chain, reflection exchanges the two central sites,
\(N/2\leftrightarrow N/2+1\). We therefore choose the seed operator for the Krylov construction to be the reflection-symmetric combination
\begin{equation}
    O
    =
    S^x_{\frac{N}{2}}
    +
    S^x_{\frac{N}{2}+1} \ .
    \label{eq:MFIM_seed_operator}
\end{equation}
Indeed, using
\begin{equation}
    \Pi S_i^x\Pi
    =
    S_{N+1-i}^x \ ,
\end{equation}
one finds
\begin{equation}
    \Pi O\Pi
    =
    S^x_{\frac{N}{2}+1}
    +
    S^x_{\frac{N}{2}}
    =
    O \ ,
    \qquad
    [O,\Pi]=0 \ .
    \label{eq:MFIM_seed_parity}
\end{equation}
Thus, the seed operator is block diagonal in the reflection-parity
basis and decomposes as
\begin{equation}
    O
    =
    O_{+}\oplus O_{-} \ ,
    \qquad
    O_q
    =
    \mathcal{P}_q O\mathcal{P}_q,
    \qquad q=\pm1 \ .
    \label{eq:MFIM_seed_blocks}
\end{equation}

For the \(N=8\) chain studied numerically, Eq.~\eqref{eq:MFIM_seed_operator}
reduces to\footnote{The Krylov recursion is initialized using the normalized operator-state associated with this seed. Therefore, an overall multiplicative normalization of \(O\) does not affect the resulting Krylov dynamics.}
\begin{equation}
    O=S_4^x+S_5^x \ .
    \label{eq:MFIM_N8_seed_operator}
\end{equation}
\begin{figure}[t!]
\centering
 \includegraphics[width=0.45\linewidth]{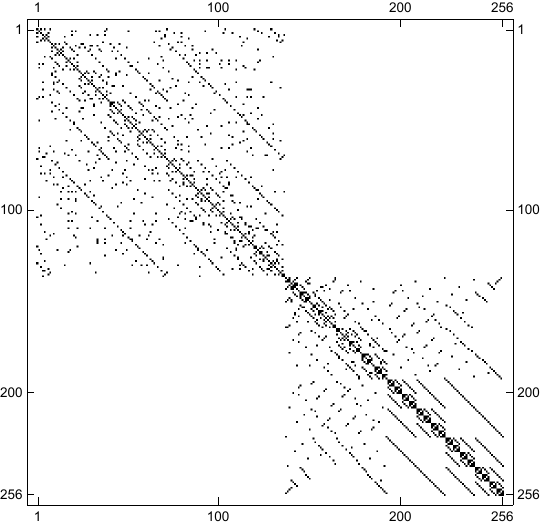}
\par\medskip
    \includegraphics[width=0.47\linewidth]{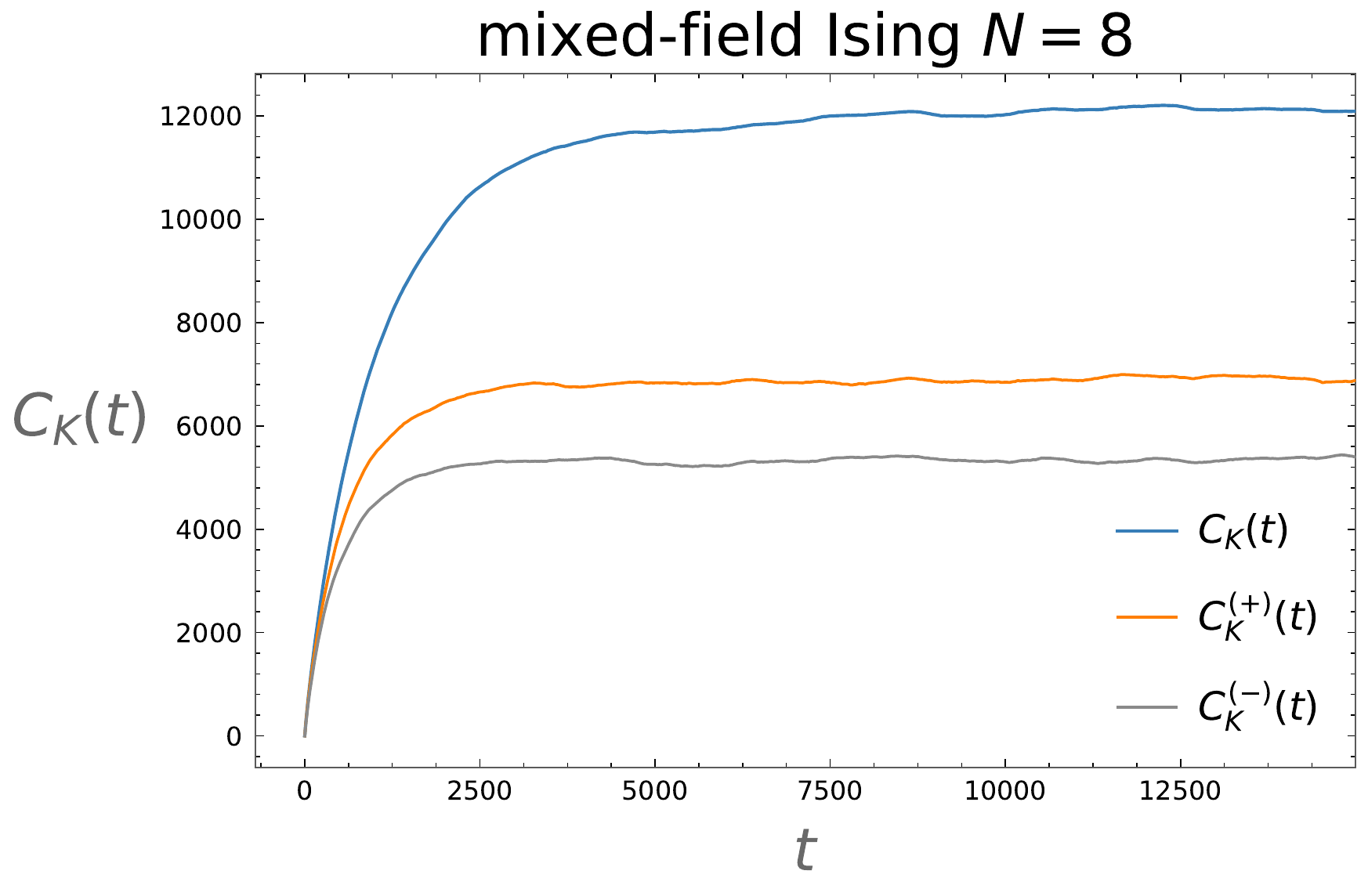}
    \hfill
    \includegraphics[width=0.51\linewidth]{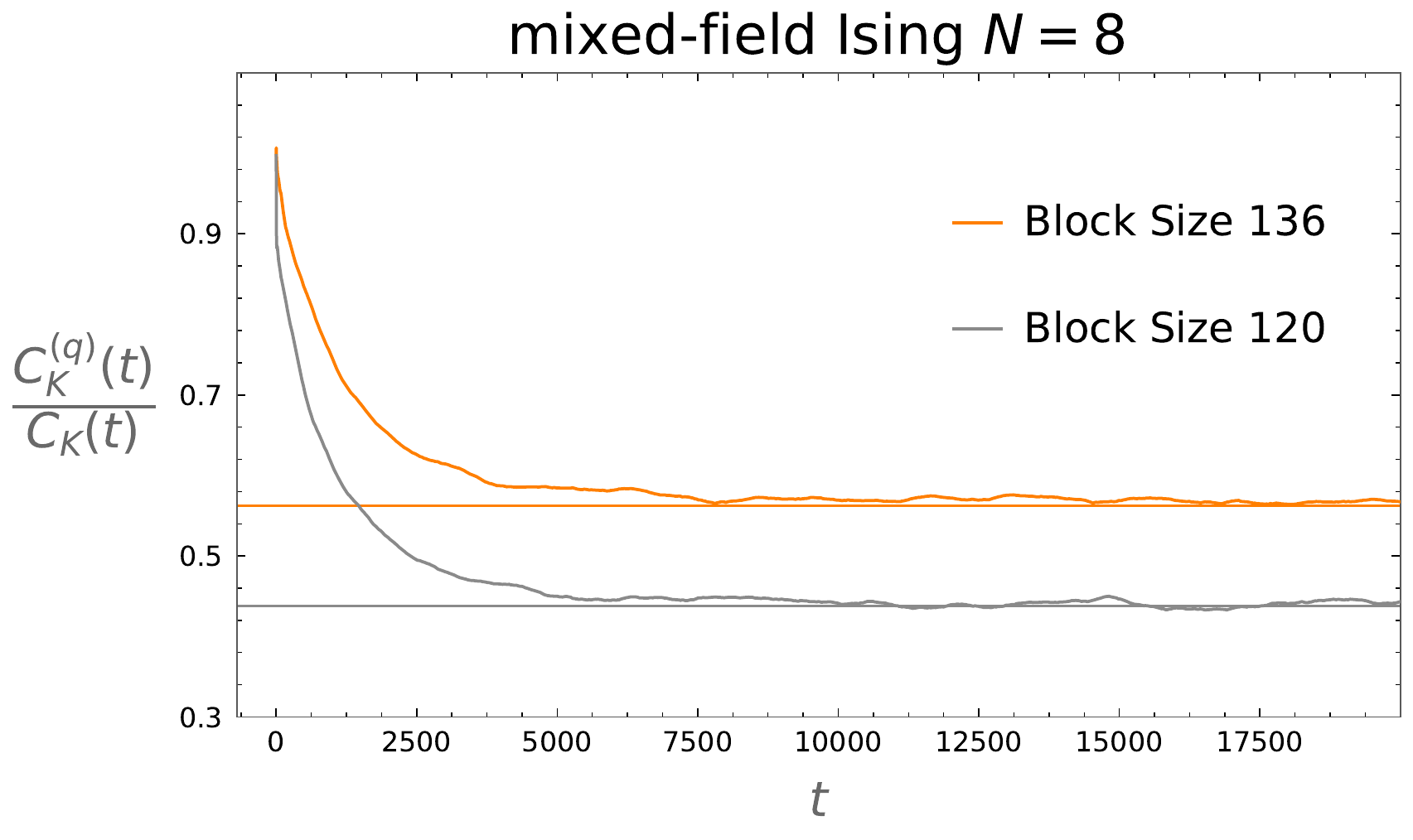}
    \caption{The upper panel shows the block-diagonal structure of the Hamiltonian in the reflection-parity basis defined by Eq.~\eqref{eq:parity_operator}. The dimensions of the reflection-even and reflection-odd sectors are \(d_{+}=136\) and \(d_{-}=120\), respectively. The bottom-left panel shows the time evolution of the full Krylov complexity \(C_K(t)\) and the symmetry-resolved complexities \(C_K^{(\pm)}(t)\) for the mixed-field Ising chain with \(N=8\). The bottom-right panel displays the corresponding fractional complexities \(C_K^{(q)}(t)/C_K(t)\), with \(q=\pm1\). The horizontal lines denote the predicted late-time values \(d_q^2/\sum_{q}d_q^2\), which agree well with the numerical results.}
\label{fig:krylov-mfi}
\end{figure}

In Fig.~\ref{fig:krylov-mfi}, we present the block-diagonal form of the Hamiltonian~\eqref{eq:MFIM_Hamiltonian} for a mixed-field Ising chain with \(N=8\) sites, together with the fractional Krylov complexities \(C_K^{(q)}(t)/C_K(t)\) for the two reflection-parity sectors. We also plot the full Krylov complexity \(C_K(t)\), together with the sector-resolved complexities \(C_K^{(+)}(t)\) and \(C_K^{(-)}(t)\), showing their complete time evolution and late-time saturation. It is evident from the figure that, at late times, the Krylov complexity is not equally distributed
between the two reflection sectors. Instead, the asymptotic contribution of each sector is governed by the dimension of its operator space, \(d_q^2\), rather than by the corresponding Hilbert-space dimension \(d_q\). For \(N=8\), the dimensions of the reflection-even and reflection-odd sectors are \(   d_{+}=136,\, d_{-}=120.\) Consequently, the predicted late-time fractions are:
\begin{equation}
    \frac{C_K^{(+)}(t)}{C_K(t)}
    \longrightarrow
    \frac{d_{+}^{\,2}}{d_{+}^{\,2}+d_{-}^{\,2}}
    =
    \frac{289}{514}
    \simeq 0.5623 \ ,
    \qquad
    \frac{C_K^{(-)}(t)}{C_K(t)}
    \longrightarrow
    \frac{d_{-}^{\,2}}{d_{+}^{\,2}+d_{-}^{\,2}}
    =
    \frac{225}{514}
    \simeq 0.4377 \ .
    \label{eq:MFIM_late_time_fractions}
\end{equation}
The numerical results approach these predicted values at late times, providing direct evidence for dimension-weighted equipartition of Krylov complexity among symmetry sectors. In contrast to the real SYK model, where the two fermion-parity sectors have equal dimensions and therefore contribute equally, the unequal reflection-sector dimensions of the
mixed-field Ising chain lead to unequal late-time contributions governed by the corresponding operator-space dimensions.

\section{Discussion}\label{sec:discussion}

In this work, we have studied the late-time behavior of Krylov complexity in quantum many-body systems with global symmetries. When the Hamiltonian and the initial operator commute with a common conserved charge, the dynamics decomposes into independent symmetry sectors, allowing one to define a Krylov complexity within each sector and compare the resulting symmetry-resolved quantities with the unresolved complexity of the full operator. Our main result is that, in chaotic finite-dimensional systems, once the Krylov wavefunction has effectively delocalized over the accessible Krylov space, the late-time complexity becomes additive over symmetry sectors,
\begin{equation}
  C_K(t\gtrsim t_{\rm sat})\simeq\sum_q C_K^{(q)}(t\gtrsim t_{\rm sat}) .
\end{equation}
Equivalently, the fractional contribution of a sector is controlled by the dimension of its dynamically accessible operator space,
\begin{equation}
  \frac{C_K^{(q)}}{C_K}
  \simeq
  \frac{d_q(d_q-1)}{\sum_{q'} d_{q'}(d_{q'}-1)}
  \simeq
  \frac{d_q^2}{\sum_{q'}d_{q'}^2}
  \qquad (d_q\gg 1),
\end{equation}
where $d_q$ is the Hilbert-space dimension of the sector labeled by $q$. The second expression is the large-sector approximation relevant for the numerical examples considered above. Thus, the late-time distribution is not governed by equal sharing among symmetry sectors, but instead exhibits a \emph{dimension-weighted equipartition}, with each sector contributing according to the size of its accessible operator space.

This behavior should be contrasted with the early-time relation discussed in Ref.~\cite{Caputa:2025mii}. At early times, the unresolved Krylov complexity is related to a weighted average of the sector-resolved complexities, with weights determined by the projection of the initial operator onto each sector. The late-time relation found here has a qualitatively different origin. After saturation, these initial-sector weights no longer control the relative contributions. Instead, effective delocalization of the Krylov wavefunction makes the late-time fractions depend only on the dimensions of the accessible Krylov spaces, up to corrections from finite-size effects, Liouvillian degeneracies, vanishing seed overlaps, and incomplete equilibration. In this sense, symmetry-resolved Krylov complexity provides a sector-wise refinement of the usual saturation diagnostic of quantum chaos.

Our numerical results support this picture across several qualitatively different systems: the real and complex SYK models, a chaotic bosonic spin model, and the mixed-field Ising chain. These examples encompass both discrete and continuous symmetries, as well as sectors of equal and unequal dimensions. In each case, the late-time plateaux of $C_K^{(q)}(t)/C_K(t)$ are well described by the dimension formula above. For equal-size sectors, the prediction reduces to an equal division among symmetry blocks. For unequal sectors, such as the charge sectors of the complex SYK model or the reflection-parity sectors of the mixed-field Ising chain, the plateaux instead track their relative operator-space dimensions, approaching the $d_q^2$ scaling for large sectors.

The present analysis relies on several assumptions whose implications deserve further investigation. First, the derivation assumes the absence of additional Liouvillian degeneracies beyond the zero modes associated with diagonal operators. Such degeneracies may arise in integrable systems, in systems with additional hidden symmetries, or at special points in parameter space, reducing the effective Krylov dimension and modifying the late-time distribution. Second, the argument assumes that the late-time Krylov wavefunction becomes approximately delocalized over the accessible Krylov space. Although this behavior is expected in chaotic systems, it need not hold in integrable or many-body localized phases. Finally, finite-size effects may be significant in exact-diagonalization studies, particularly for small spin chains where the separation between saturation and recurrence time scales is limited \cite{Capizzi:2025eyg}.

Several extensions would be interesting to pursue. A natural first step is to characterize the approach to the late-time plateau and determine the leading finite-time and finite-size corrections to the dimension-weighted equipartition rule. It would also be useful to study systems with non-Abelian symmetries, where representation multiplicities lead to a richer sector structure and the appropriate notion of symmetry-resolved operator growth may be more intricate. Another direction is to investigate open quantum systems, Floquet dynamics, and systems with weakly broken symmetries, where the block structure is either modified or only approximately preserved. More broadly, it would be interesting to clarify the relation between dimension-weighted Krylov equipartition, random-matrix universality within symmetry sectors, and other symmetry-resolved probes of quantum chaos.

Overall, our results show that resolving symmetries is essential for a precise characterization of operator growth at late times. The saturation value of Krylov complexity is controlled not by the square of the total Hilbert-space dimension, but by the sum of the dimensions of the dynamically accessible operator spaces, which reduces to $\sum_q d_q^2$ for large sectors. Symmetry resolution therefore removes an important ambiguity in the interpretation of late-time Krylov complexity and provides a sharper characterization of chaotic operator dynamics.

\section*{Acknowledgements}
We thank Kyoung-Bum Huh and Hyun-Sik Jeong for their initial collaboration, and Arnab Kundu for useful discussions. SD, JFP and LCQ are supported by the ‘Atracción de Talento’ program of the Comunidad de Madrid under grant 2020-T1/TIC-20495, the Spanish Agencia Estatal de Investigación through grants CEX2025-001574-S, PID2021-123017NB-I00 and PID2024-156043NB-I00, funded by MCIN/AEI/10.13039/501100011033, and ERDF, EU. LCQ further acknowledges support from the Chinese Scholarship Council (CSC) through a graduate scholarship. The research presented in this publication falls within the research line Strings and Quantum Gravity

\appendix

\section{Late-time saturation of the Krylov wavefunction}
\label{app:wavefn_property}

\begin{figure}[t!]
\centering
    \includegraphics[width=0.49\linewidth]{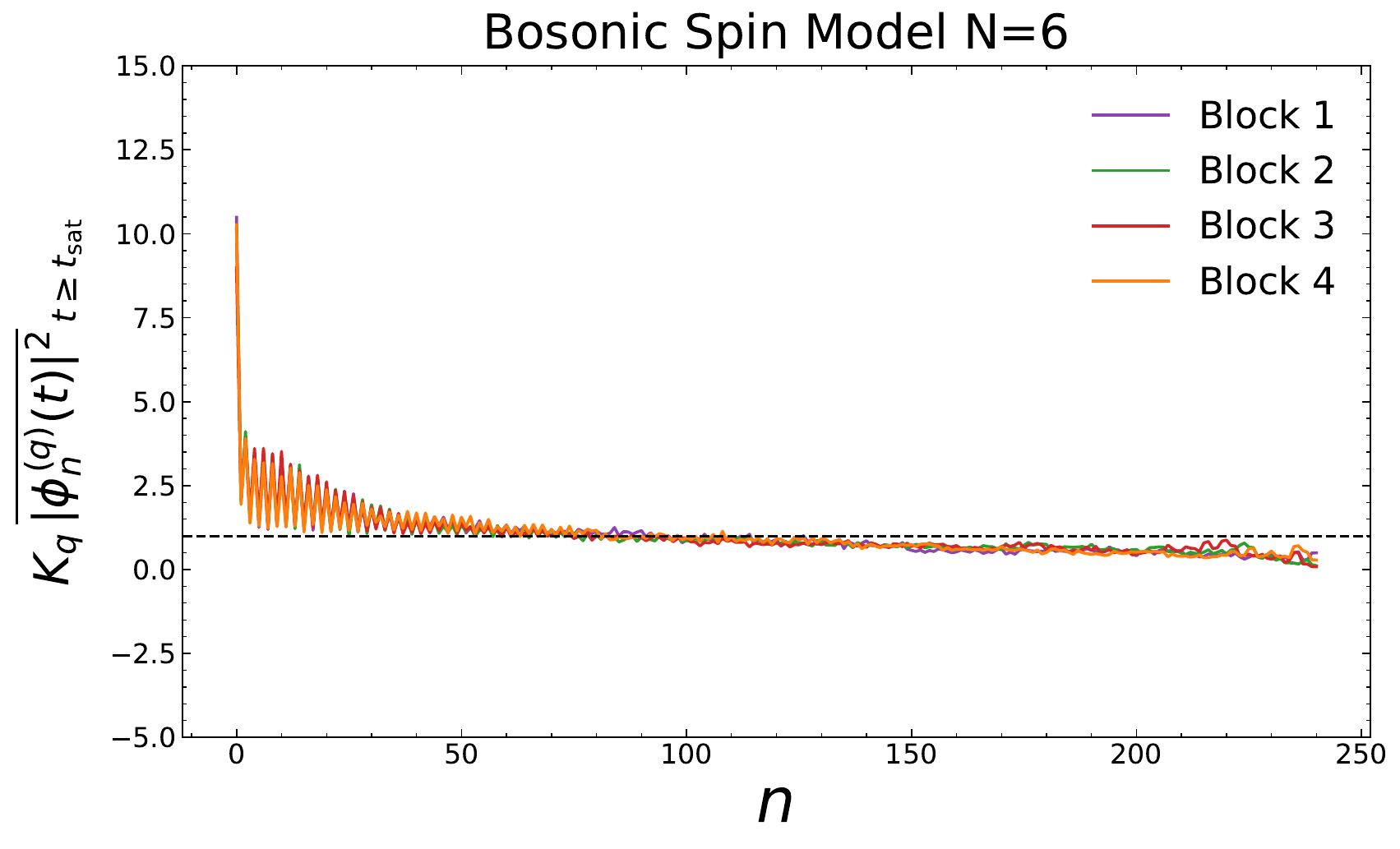}
    \hfill
    \includegraphics[width=0.49\linewidth]{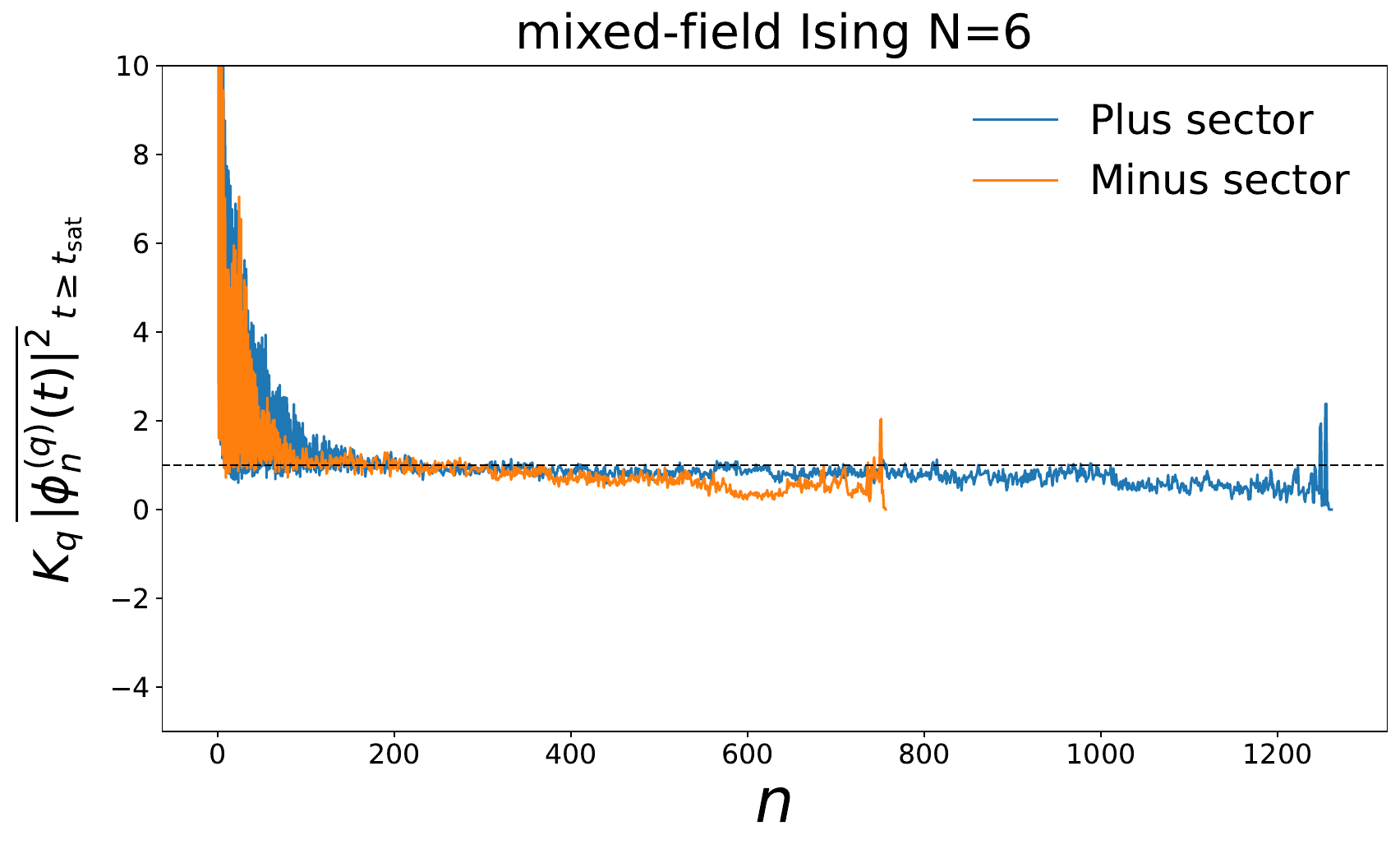}
    \caption{Late-time rescaled probability distributions
\(K_q\,\overline{|\phi_n^{(q)}(t)|^2}_{\,t\geq t_{\rm sat}}\)
for each symmetry sector of the chaotic bosonic spin model (left) and the chaotic mixed-field Ising model (right), both for \(N=6\), where \(K_q\) denotes the corresponding sector Krylov dimension. The horizontal dashed line at unity represents the uniform-distribution expectation
\(\overline{|\phi_n^{(q)}(t)|^2} \sim\cfrac{1}{K_q}\), while the probability distribution satisfies the normalization condition \(\sum_{n=0}^{K_q-1}
\overline{|\phi_n^{(q)}(t)|^2}_{\,t\geq t_{\rm sat}}=1\).}
\label{fig:krylov_wavefunction}
\end{figure}

In section~\ref{subsec:late_time_Krylov}, we derived an analytic expression for the symmetry-resolved Krylov complexity $C_K^{(q)}$ under the assumption that, beyond the saturation time $t_{\rm sat}$, the Krylov wavefunction becomes approximately delocalized over the accessible Krylov chain:
\begin{equation}
    \left|\phi_n^{(q)}(t)\right|^2
    \simeq
    \frac{1}{K_q},
    \qquad
    t\gtrsim t_{\rm sat},
    \qquad
    n=0,\ldots,K_q-1.
\end{equation}
In this appendix, we verify this assumption numerically. Fig.~\ref{fig:krylov_wavefunction} shows the late-time-averaged Krylov probability distribution,
$\overline{\left|\phi_n^{(q)}(t)\right|^2}$, as a function of $n$ for the bosonic spin model and the mixed-field Ising model. In both cases, the distribution is approximately uniform along the Krylov chain, confirming the validity of the above assumption at late times.
%

\section{Symmetry-resolved Krylov complexity in integrable models}
\label{app:Krylov_integrable}
\raggedbottom

In the main text, we showed that chaotic systems exhibit the
approximate late-time additivity
\begin{equation}
    C_K(t\gtrsim t_{\mathrm{sat}})
    \simeq
    \sum_q C_K^{(q)}(t\gtrsim t_{\mathrm{sat}}),
\end{equation}
which implies that the late-time fractional Krylov complexities are governed by
\begin{equation}
    \frac{C_K^{(q)}}{C_K}
    \simeq
    \frac{d_q(d_q-1)}
    {\displaystyle\sum_{q'}d_{q'}(d_{q'}-1)}
    \simeq
    \frac{d_q^2}{\displaystyle\sum_{q'}d_{q'}^2},
\end{equation}
This relies on the fact that the sector Krylov dimensions saturate their generic bounds and that no additional cross-sector degeneracies are present:
\begin{equation}
    K_q=d_q(d_q-1)+1,
    \qquad
    \Omega_q\cap\Omega_{q'}=\{0\},
    \qquad
    q\neq q'.
\end{equation}

In an integrable system, degeneracies of the Liouvillian frequencies and selection rules associated with the seed can substantially reduce the true Krylov dimensions below \(d_q(d_q-1)+1\). The example studied below exhibits precisely this behavior: the two reflection-sector frequency sets remain disjoint, while the true Krylov dimensions are far below the corresponding generic chaotic bounds. As we show numerically, the late-time fractional complexities then saturate near the values set by the relative dimensions of the true seed-accessible Krylov spaces, rather than the \(d_q^2\)-weighted prediction.\footnote{As noted for
the \(N=14\) real SYK model below Eq.~(3.18), an additional unitary or antiunitary symmetry can relate distinct symmetry sectors and render them isospectral. A similar mechanism may occur in other models, with the seed accessing the same Liouvillian frequencies in more than one sector. These shared frequencies are counted only once in
\(K_{\mathrm{full}}=\left|\bigcup_q\Omega_q\right|\), whereas they are counted separately in the independently constructed sector Krylov spaces. The dimension relation underlying late-time additivity may therefore fail. Strictly, isospectrality of the Hamiltonian blocks alone is not sufficient; their seed-accessible Liouvillian frequency sets must overlap.}

A direct operator-space Arnoldi recursion can become numerically unstable when the true Krylov space occupies only a restricted subspace of the full operator space. Finite-precision errors may introduce components outside the exact cyclic subspace, which are subsequently normalized and interpreted as additional Krylov directions. The recursion can therefore deviate from the exact coefficients and continue beyond the true endpoint.

We first diagonalize the Hamiltonian and determine the distinct Liouvillian frequencies having nonzero spectral weight. Numerically degenerate frequencies are grouped using an adaptive absolute tolerance. Specifically, we scan $\epsilon_{\omega}\in\left\{10^{-13},10^{-12},\ldots,10^{-6}\right\}$ and record the resulting number of distinct-frequency groups. We identify the longest plateau over which this number remains unchanged and choose its central value as the frequency-grouping tolerance. After sorting the frequencies, consecutive values satisfying $\omega_{\alpha+1}-\omega_{\alpha}\leq \epsilon_{\omega}$ are treated as numerically identical, and their spectral weights are added. Let \(\{\omega_{\alpha}\}_{\alpha=1}^{K}\) denote the resulting distinct seed-accessible frequencies, with corresponding weights \(W_{\alpha}\equiv W(\omega_{\alpha})\). We then introduce the diagonal matrix
\begin{equation}
    D=
    \operatorname{diag}
    \left(
        \omega_1,\omega_2,\ldots,\omega_K
    \right),
\end{equation}
together with the normalized initial vector
\begin{equation}
    |V_0)
    =
    \frac{1}{\sqrt{\sum_{\alpha=1}^{K}W_{\alpha}}}
    \begin{pmatrix}
        \sqrt{W_1}\\
        \sqrt{W_2}\\
        \vdots\\
        \sqrt{W_K}
    \end{pmatrix}.
\end{equation}
The pair $(D,|V_0))$ has the same spectral measure as the original pair $(\mathcal{L},|O_0))$ and therefore generates the same Lanczos coefficients within the seed-accessible Krylov space. We may thus apply the standard Lanczos--Arnoldi recursion given in Eq.~\eqref{eq:arnoldi_alogo} directly to
$(D,|V_0))$. In the numerical implementation, we use the two re-orthogonalization passes specified in Eq.~\eqref{eq:arnoldi_alogo}. The recursion is terminated when
\begin{equation}
    h_{k,k-1}
    <
    \epsilon_{\mathrm{A}}\,h_{\max}^{(k)},
    \qquad
    \epsilon_{\mathrm{A}}=10^{-8},
    \qquad
    h_{\max}^{(k)}
    =
    \max_{1\leq j\leq k}h_{j,j-1},
\end{equation}
or once all $K$ seed-accessible frequency directions have been exhausted.

\begin{figure}[t!]
\centering

\includegraphics[width=0.6\textwidth]{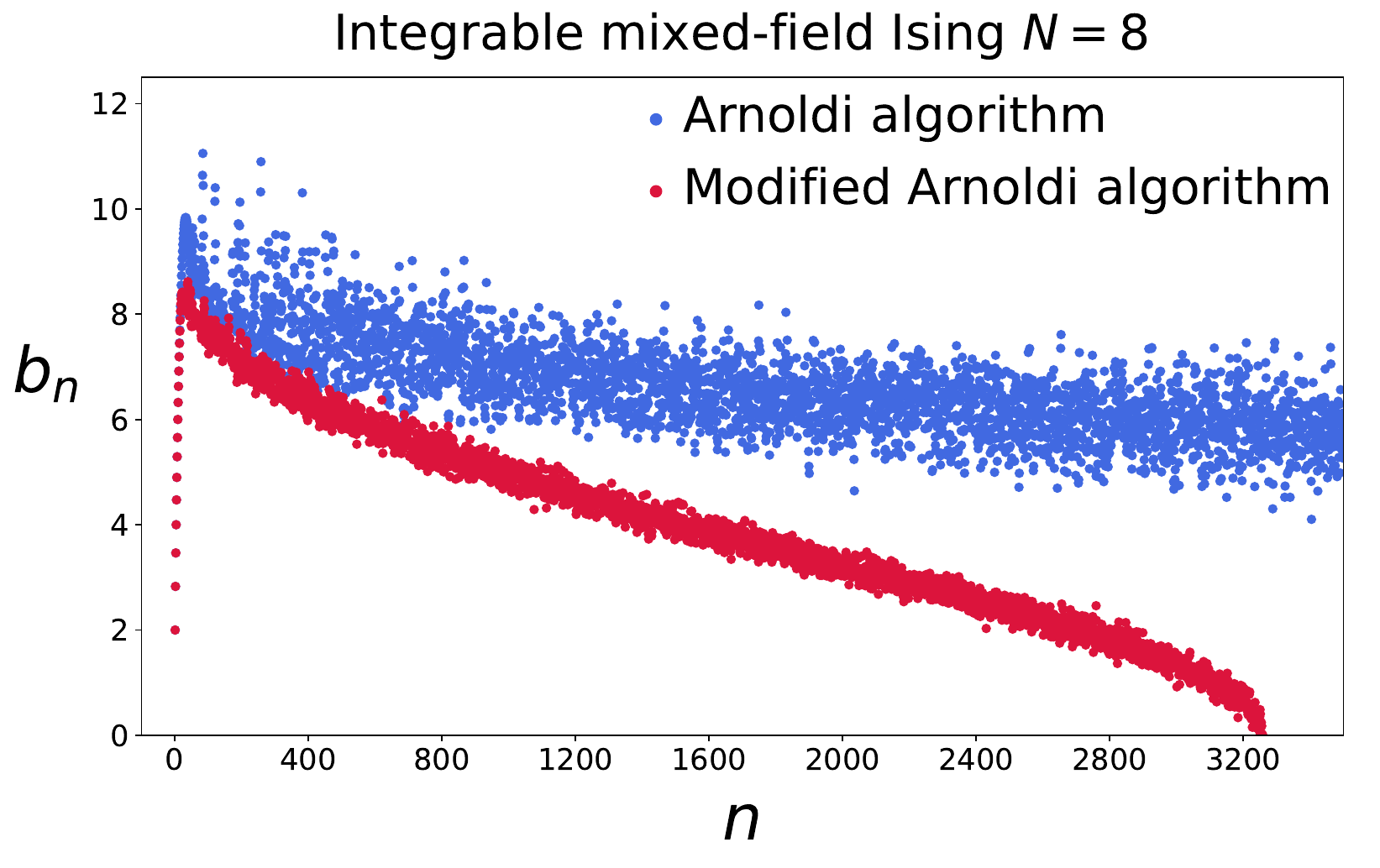}
\caption{Comparison of the Lanczos coefficients \(b_n\) obtained from the operator-space Arnoldi recursion and the modified Lanczos construction for the integrable mixed-field Ising chain with \(N=8\). The modified algorithm correctly terminates at the true full Krylov dimension \ \(K_{\mathrm{full}}=3258\), whereas the ordinary algorithm generates spurious nonzero coefficients beyond this dimension.}
\label{fig:integrable_algorithm_comparison}
\end{figure}

Fig.~\ref{fig:integrable_algorithm_comparison} compares the
operator-space Arnoldi recursion with the spectral-measure Lanczos construction for the integrable mixed-field Ising chain. The operator-space calculation becomes numerically unstable and continues beyond the true endpoint, whereas the spectral construction correctly terminates at \(K_{\mathrm{full}}=3258\). We therefore use the
spectral-measure construction for the results presented in this appendix.\footnote{For the chaotic models considered in the main text, we have verified that both algorithms yield the same Lanczos coefficients $b_n$ and Krylov dimension.}

With this numerical framework in place, we now turn to the integrable transverse-field Ising chain, where the reflection-sector frequency sets are disjoint but selection rules strongly reduce the true Krylov dimensions below their generic chaotic bounds.

The integrable transverse-field limit of the mixed-field Ising chain~\eqref{eq:MFIM_Hamiltonian}, obtained by setting
\begin{equation}
    h_x=-1,
    \qquad
    h_z=0.
\end{equation}
The Hamiltonian then becomes
\begin{equation}
H = - \sum_{i=1}^{N-1} S_i^z S_{i+1}^z 
+ \sum_{i=1}^{N} S_i^x \ ,
\label{eq:MFIM_Integrable__Hamiltonian}
\end{equation}
It preserves the reflection symmetry \(\Pi\). \footnote{At \(h_z=0\), the Hamiltonian acquires the global spin-flip symmetry
\(\Gamma_x=\prod_{i=1}^{N}S_i^x\), under which the chosen seed is odd, \(\Gamma_x O\Gamma_x=-O\). Consequently, in a simultaneous eigenbasis of \(H\) and \(\Gamma_x\), the diagonal matrix elements of the seed vanish. For the finite-size spectrum considered here, this implies
\(W_\pm(0)=0\), consistent with the numerically observed disjointness
\(\Omega_+\cap\Omega_-=\varnothing\).} For even \(N\), we choose the reflection-symmetric seed operator
\begin{equation}
    O
    =
    S_{N/2}^{z}+S_{N/2+1}^{z}.
\end{equation}
For \(N=8\), the operator is 
\begin{equation}
    O=S_4^z+S_5^z,
    \qquad
    d_{+}=136,
    \qquad
    d_{-}=120.
\end{equation}
Using the spectral counting described above, in which only distinct
frequencies with nonzero seed matrix elements are retained, we obtain
\begin{equation}
    K_{+}=2028,
    \qquad
    K_{-}=1230.
\end{equation}
Within numerical precision, the two accessible-frequency sets are disjoint, \(\Omega_{+}\cap\Omega_{-}=\varnothing\), and therefore
\begin{equation}
    K_{\mathrm{full}}
    =
    |\Omega_{+}\cup\Omega_{-}|
    =
    K_{+}+K_{-}
    =
    3258.
\end{equation}

These values are far below the generic chaotic bounds,
\begin{align}
    K^{+}_{\mathrm{max}}
    &=d_{+}(d_{+}-1)+1=18361,
    \\
    K^{-}_{\mathrm{max}}
    &=d_{-}(d_{-}-1)+1=14281,
    \\
    K_{\mathrm{max}}
    &=1+\sum_{q=\pm}d_q(d_q-1)=32641.
\end{align}
In particular, the seed accesses only a small fraction of the available operator space, showing that the chaotic scaling \(K_q\simeq d_q^2\) does not hold in this integrable limit.

\begin{figure}[t!]
\centering
\includegraphics[width=0.49\linewidth]{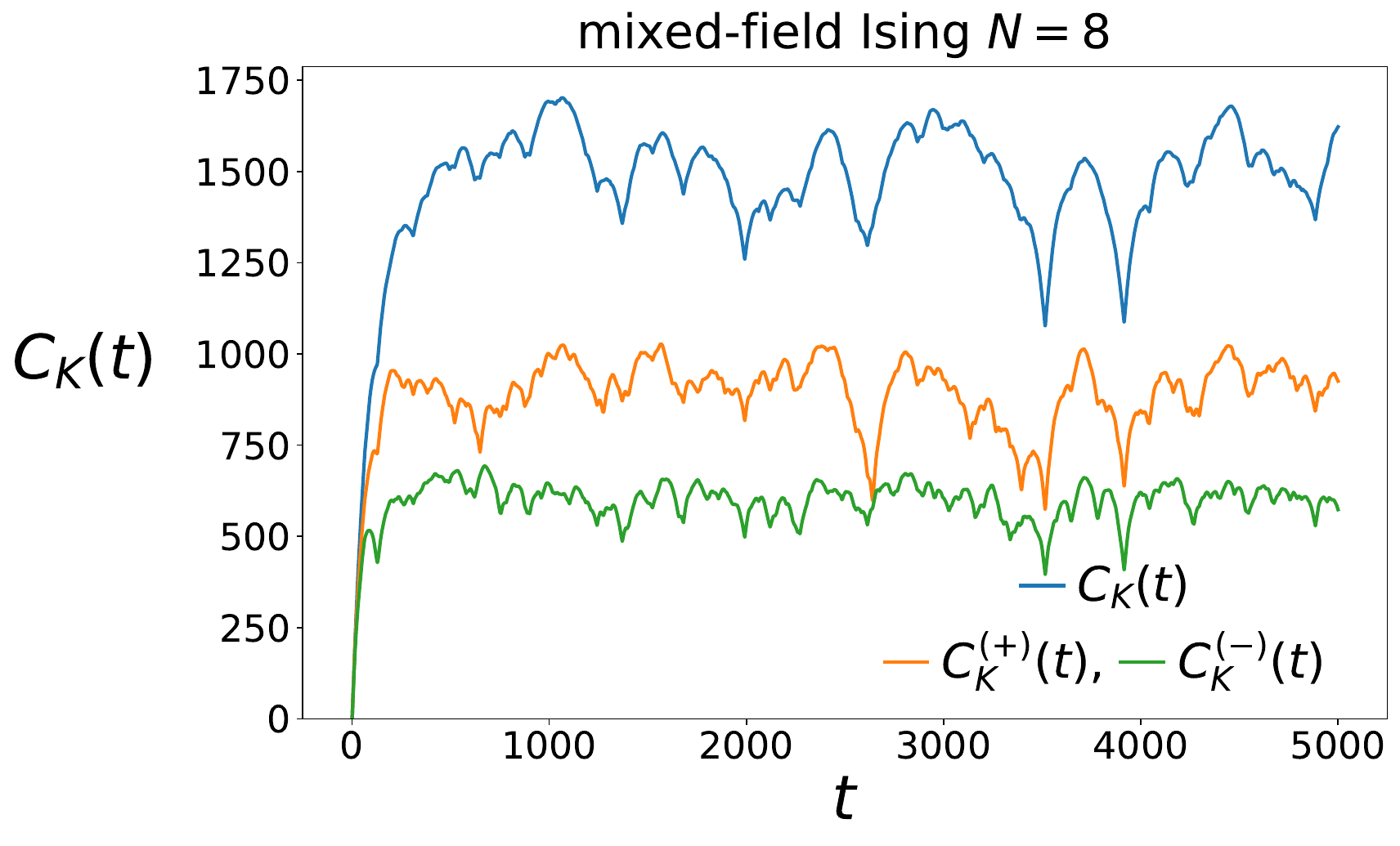}
\hfill
\includegraphics[width=0.49\linewidth]{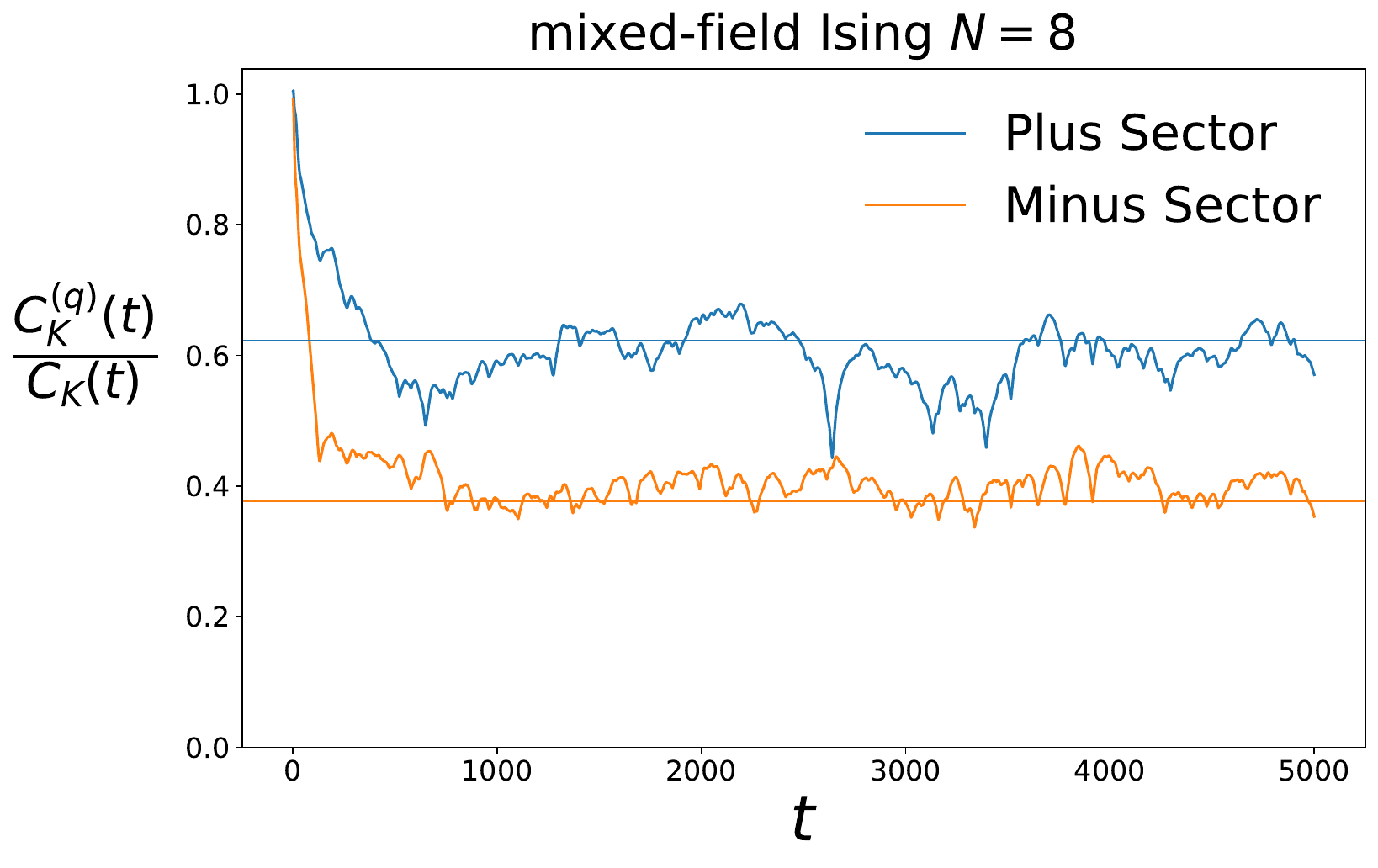}
\caption{The left panel shows the time evolution of the full Krylov complexity \(C_K(t)\) and the reflection-resolved complexities \(C_K^{(q,+)}(t)\) and \(C_K^{(q,-)}(t)\) for the mixed-field Ising chain with \(N=8\) in the integrable transverse-field limit, \(h_x=-1\) and \(h_z=0\). The right panel displays the corresponding fractional complexities \(C_K^{(q,\pm)}(t)/C_K(t)\). The horizontal straight lines denote the predictions \(\frac{K_q-1}{K_{\mathrm{full}}-1},\) obtained from the true Krylov dimensions \(K_{+}=2028\), \(K_{-}=1230\), and \(K_{\mathrm{full}}=3258\), giving approximately \(0.6224\) and \(0.3773\) for the reflection-even and reflection-odd sectors, respectively.}
\label{fig:krylov_integrable_mfi}
\end{figure}

Fig.~\ref{fig:krylov_integrable_mfi} illustrates a complementary integrable
case in which the sector frequency sets are disjoint, while the true
Krylov dimensions remain substantially below their generic chaotic
bounds. Consequently, the late-time fractional Krylov complexities do
not approach the dimension-weighted values
\(d_q^2/\sum_{q'}d_{q'}^2\), but instead saturate near
\((K_q-1)/(K_{\mathrm{full}}-1)\), as determined by the relative
dimensions of the true seed-accessible Krylov spaces.

\bibliography{bibliography}
\bibliographystyle{JHEP.bst}

\end{document}